\documentclass[traditabstract, longauth]{aa} 
\usepackage{graphicx}
\usepackage{subfigure}
\usepackage{multirow}

\usepackage{txfonts}
\usepackage{natbib}
\bibpunct{(}{)}{;}{a}{}{,} 
\begin{document}
   \title{X-shooter, the new wide band intermediate resolution
     spectrograph at the ESO Very Large Telescope}

\author{J. Vernet \inst{1}  \and 
H.    Dekker \inst{1}   \and 
S.    D'Odorico \inst{1}  \and 
L.    Kaper \inst{2} \and 
P.    Kjaergaard \inst{3} \and 
F.    Hammer \inst{4} \and 
S.    Randich \inst{5} \and 
F.    Zerbi \inst{6} \and 
P. M. Groot   \inst{7} \and 
J.    Hjorth \inst{3} \and
I.    Guinouard \inst{4} \and 
R.    Navarro \inst{8} \and 
T.    Adolfse \inst{7} \and 
P. W. Albers  \inst{7} \and 
J.-P. Amans \inst{4} \and 
J. J. Andersen \inst{3} \and 
M. I. Andersen \inst{3} \and 
P.    Binetruy \inst{9} \and 
P.    Bristow \inst{1}   \and 
R.    Castillo \inst{10} \and 
F.    Chemla \inst{4} \and 
L.    Christensen \inst{11} \and 
P.    Conconi  \inst{6} \and 
R.    Conzelmann \inst{1}   \and 
J.    Dam \inst{7} \and 
V.    De Caprio  \inst{12} \and 
A.    De Ugarte Postigo \inst{3} \and 
B.    Delabre \inst{1}   \and 
P.    Di Marcantonio  \inst{13} \and 
M.    Downing \inst{1}   \and 
E.    Elswijk \inst{8} \and 
G.    Finger \inst{1}   \and 
G.    Fischer \inst{1} \and 
H.    Flores \inst{4} \and 
P.    Fran\c{c}ois \inst{4} \and 
P.    Goldoni \inst{9} \and 
L.    Guglielmi \inst{9} \and 
R.    Haigron \inst{4} \and 
H.    Hanenburg \inst{8} \and 
I.    Hendriks\inst{7} \and 
M.    Horrobin \inst{14} \and 
D.    Horville \inst{4} \and 
N. C. Jessen  \inst{15} \and   
F.    Kerber \inst{1}   \and 
L.    Kern \inst{1} \and 
M.    Kiekebusch \inst{1} \and 
P.    Kleszcz \inst{8} \and 
J.    Klougart   \inst{3} \and 
J.    Kragt \inst{8} \and 
H. H. Larsen \inst{3} \and 
J.-L. Lizon \inst{1}  \and 
C.    Lucuix \inst{1}   \and 
V.    Mainieri \inst{1} \and 
R.    Manuputy \inst{16} \and 
C.    Martayan \inst{10} \and 
E.    Mason \inst{17} \and 
R.    Mazzoleni \inst{6}   \and 
N.    Michaelsen \inst{3} \and 
A.    Modigliani \inst{1}   \and 
S.    Moehler \inst{1} \and 
P.    M{\o}ller \inst{1} \and 
A.    Norup S{\o}rensen \inst{3} \and 
P.    N{\o}rregaard \inst{3} \and 
C.    P\'{e}roux \inst{18}   \and 
F.    Patat \inst{1} \and 
E.    Pena \inst{10} \and 
J.    Pragt \inst{8} \and 
C.    Reinero \inst{10} \and 
F.    Rigal \inst{8} \and 
M.    Riva  \inst{6} \and 
R.    Roelfsema \inst{8} \and 
F.    Royer \inst{4} \and 
G.    Sacco   \inst{19} \and 
P.    Santin  \inst{13} \and 
T.    Schoenmaker \inst{8} \and 
P.    Spano  \inst{6} \and 
E.    Sweers \inst{7} \and 
R.    Ter Horst \inst{8} \and 
M.    Tintori  \inst{20} \and 
N.    Tromp \inst{8} \and 
P.    van Dael  \inst{7} \and 
H.    van der Vliet \inst{7} \and 
L.    Venema \inst{8} \and 
M.    Vidali  \inst{21} \and 
J.    Vinther \inst{1}   \and 
P.    Vola \inst{18} \and 
R.    Winters \inst{7} \and 
D.    Wistisen \inst{3} \and 
G.    Wulterkens  \inst{7} \and 
A.    Zacchei  \inst{13} 
}

   \institute{European Southern Observatory, Karl Schwarzschild Strasse 2, D-85748 Garching bei M\"unchen, Germany 
        \and
        Astronomical Institute ‘Anton Pannekoek’, University of Amsterdam, Kruislaan 403, 1098 SJ Amsterdam, the Netherlands
        \and
         Niels Bohr Institute, Juliane Maries Vej 30, DK-2100 Copenhagen, Denmark
        \and 
         GEPI - Observatoire de Paris, 5 place Jules Janssen, 92195 Meudon, France
        \and
         INAF - Osservatorio Astrofisico di Arcetri, Largo E. Fermi 5, 50125 Firenze, Italia 
        \and
         INAF - Osservatorio Astronomico di Brera, Via E. Bianchi 46, 23807 Merate, Italy 
         \and
        Radboud Univ. Nijmegen, Postbus 9010, 6500 GL Nijmegen, The Netherlands
        \and
         ASTRON, Oude Hoogeveensedijk 4, 7991 PD Dwingeloo, The Netherands
        \and
        Laboratoire Astroparticule et Cosmologie, 10 rue A. Domon et L. Duquet, 75205 Paris Cedex 13, France
        \and
         European Southern Observatory, Alonso de C\'{o}rdova 3107
        Vitacura, Casilla 19001 Santiago de Chile 19, Chile
        \and
         Excellence Universe Cluster, Technische Universit\"at M\"unchen,
        Bolzmannstr. 2, 85748 Garching bei M\"unchen, Germany
        \and
         INAF - Osservatorio Astronomico di Capodimonte, Salita
        Moiariello 16, 80131 Napoli
        \and
         INAF - Osservatorio Astronomico di Trieste, Via Tiepolo 11, 34143 Trieste, Italy 
         \and
         Physikalisches Institut Universit\"at zu K\"oln, Z\"ulpicher
        Strasse 77, 50937 K\"oln 
        \and
        DTU Space, Juliane Maries vej 30, DK-2100, Copenhagen, Denmark
        \and
        Technologie Centrum FNWI, Science Park 904, 1098 XH
        Amsterdam, The Netherlands 
        \and
         Space Telescope Science Institute, I3700 San Martin Drive, Baltimore, MD 21218, USA 
        \and
         Laboratoire d’Astrophysique de Marseille, OAMP, 
         Universit\'{e} Aix-Marseille \& CNRS, 38 rue Fr\'{e}d\'{e}ric Joliot Curie, 13388 Marseille c\'{e}dex 13, France 
        \and
         Chester F. Carlson Center for Imaging Science, Rochester Institute of Technology, 54 Lomb Memorial Drive, 14623 Rochester,
USA
         \and
         A.D.S. International S.r.l., via Roma 87, 23868 Valmadrera,
        Italy
        \and
         EC JRC Institute for Reference Materials and Measurements, B-2440 Geel, Belgium 
             }

   \date{Received July 21, 2011; accepted September 27, 2011}

 
  \abstract{

    X-shooter is the first $2^{\rm nd}$ generation instrument of the
    ESO {\it Very Large Telescope} (VLT). It is a very efficient,
    single-target, intermediate-resolution spectrograph that was installed at
    the Cassegrain focus of UT2 in 2009. The instrument covers, in a
    single exposure, the spectral range from 300 to 2500~nm. It is
    designed to maximize the sensitivity in this spectral range
    through dichroic splitting in three arms with optimized optics,
    coatings, dispersive elements and detectors. It operates at
    intermediate spectral resolution ($R \sim 4,000 - 17,000$,
    depending on wavelength and slit width) with fixed \'echelle
    spectral format (prism cross-dispersers) in the three arms. It
    includes a 1.8\arcsec$\times$4\arcsec~Integral Field Unit as an
    alternative to the 11\arcsec~long slits. A dedicated data
    reduction package delivers fully calibrated two-dimensional and
    extracted spectra over the full wavelength range. We describe the
    main characteristics of the instrument and present its performance
    as measured during commissioning, science verification and the
    first months of science operations.

}


   \maketitle
%

\section{Introduction}

On November 19, 2001, the European Southern Observatory (ESO) issued a
Call for Proposals for $2^{\rm nd}$ generation VLT instruments. Four
instrument concepts were quoted in the Call as being of particular
interest to the ESO community: a cryogenic multi-object spectrometer
in the 1 to 2.4~$\mu$m range, a wide-field 3D optical spectrometer, a
high contrast, adaptive optics assisted, imager (Planet Finder) and a
medium resolution, wide-band (0.32--2.4 $\mu$m) spectrometer. In 2002
four proposals were received on this latter instrument
concept. Following a phase of discussions between the various
proponents and ESO, a consortium was formed between ESO and several
partner institutes in Denmark, France, Italy and the Netherlands to
carry out a Feasibility Study. After interaction with the ESO
Scientific and Technical Committee (STC), the study was presented in final
form to ESO in October 2003 and the project was finally approved by
the ESO Council in December 2003 \citep{Dodorico2004,Dodorico2006}.

After a Preliminary Design Review in 2004 and a Final Design Review in
2006, the different hard- and software components of the instrument were
manufactured, integrated and tested at the 10 consortium institutes
(Table~\ref{contributions}). Final integration of the full instrument
was done at ESO Headquarters in Garching, Germany, before it was
installed and commissioned on the VLT at ESO Paranal, Chile
\citep{vernet09}. The instrument was completed in 5 years
(along with the near-IR camera HAWK-I, the
shortest construction time of VLT instruments so far) at a cost of
about 5.3 million Euro and about 70 person-years. The major fraction of
the hardware costs and manpower were provided by the consortium
institutes, and compensated by ESO with Guaranteed Time Observations
(GTO) on the VLT. The X-shooter GTO program includes about 150 nights
and is being executed in a three-year period (2009--2012).

\begin{table*}
\caption{\label{contributions} An overview of the hardware and software contributions to
X-shooter from the different consortium partners.}
\begin{center}
\begin{tabular}{l l} 
\hline
\hline
\textbf{ESO} & \textbf{PI S. D'Odorico, PM H. Dekker} \\

\hline
    & Detector systems \\
    & Flexure compensation system \\
    & Cryogenic control electronics\\
    & Data reduction software \\ 
    & Final integration and commissioning \\ 
\hline
\textbf{Denmark} & \textbf{PI and PM P. Kjaergaard Rasmussen} \\ 
\hline
Niels Bohr Institute, Copenhagen University, & UVB \& VIS spectrograph (mechanics)
\\
DTU Space, Copenhagen & Instrument backbone \\
                             & Pre-slit optics and calibration system
                             \\
                             & Instrument control electronics \\
                             \hline 
\textbf{France} & \textbf{PI F. Hammer, PM I. Guinouard} \\ 
\hline
Paris Observatory Meudon & Integral field unit \\
Astroparticle and Cosmology Paris& Data reduction software \\ 
\hline
\textbf{Italy} & \textbf{PI R. Pallavicini/S. Randich, PM F. Zerbi} \\ 
\hline
INAF Observatory Palermo, & UVB \& VIS spectrographs (optics) \\
INAF Observatory Brera, & Instrument Control Software \\
INAF Observatory Trieste, &  \\
INAF Observatory Catania &  \\
 \hline
\textbf{The Netherlands} & \textbf{PI L. Kaper, PM R. Navarro} \\
 \hline
Netherlands Research School for Astronomy (NOVA) & NIR spectrograph \\
Astronomical Institute, University of Amsterdam & NIR cryostat \\
Astronomical Institute, Radboud University Nijmegen & Data reduction
software \\ 
ASTRON Dwingeloo & \\
\hline
\end{tabular}
\end{center}
\end{table*}

The X-shooter project was led by a Project Board consisting of
four national Principal Investigators (PIs) and the ESO PI (see
Table~\ref{contributions}) and met about once every six months. Sadly,
the Italian co-PI R. Pallavicini passed away on January 10, 2009. The
overall project management and system engineering was in hands of
H. Dekker (ESO), assisted by F. Zerbi (ESO). The X-shooter Science
Team was led by J. Hjorth (DK) and subsequently P.M. Groot
(NL). J. Vernet (ESO) has been appointed as the X-shooter Instrument
Scientist. Activities within each country were coordinated by a national
project manager (PM); the work package managers reported to their
national PM who had a monthly teleconference with the other PMs and
H. Dekker.

The concept of X-shooter has been defined with one principle goal in
mind: the highest possible throughput over the wavelength range from
the atmospheric cutoff to the near infrared at a resolution where the
instrument is sky limited in a half hour exposure. Sky
background and detector noise considerations require that the spectral
resolution is in the range 5,000 to 7,000 (for a 1\arcsec slit). This
resolution also ensures that in the near infrared, 80 to 90\% of the
detector pixels are not affected by strong sky lines so that most of
the covered spectrum is sky background continuum limited.
Other constraints on the design of X-shooter were set by the size and
weight (and related flexure) limits of the VLT Cassegrain focus.

The X-shooter Science Case (see ESO/STC-324A
is broad and includes various applications
ranging from nearby intrinsically faint stars to bright sources at the
edge of the Universe. X-shooter's unique wavelength coverage and high
efficiency opens a new observing capacity in observational
astronomy. Key science cases to be addressed with X-shooter concern
the study of brown dwarfs, young stellar objects and T Tauri stars,
the progenitors of supernovae Type Ia, gamma-ray bursts (GRB), quasar
absorption lines, and lensed high-z galaxies. The advantage of the
large wavelength coverage is that e.g. the redshift of the target does
not need to be known in advance (as is the case for GRBs); also,
the study of Lyman~$\alpha$ in high-redshift galaxies will be possible
in the redshift range $1.5 < z < 15$.  VLT/X-shooter will complement
and benefit from other major facilities in observational astrophysics
operational in the period 2010--2020: survey instruments like
VST/OmegaCAM and VISTA working in the same wavelength range, and
observatories like LOFAR, ALMA, JWST, {\it Swift} and {\it Fermi}
exploring other observing windows.

\begin{figure}
 \begin{center}
\includegraphics[height=5.6cm]{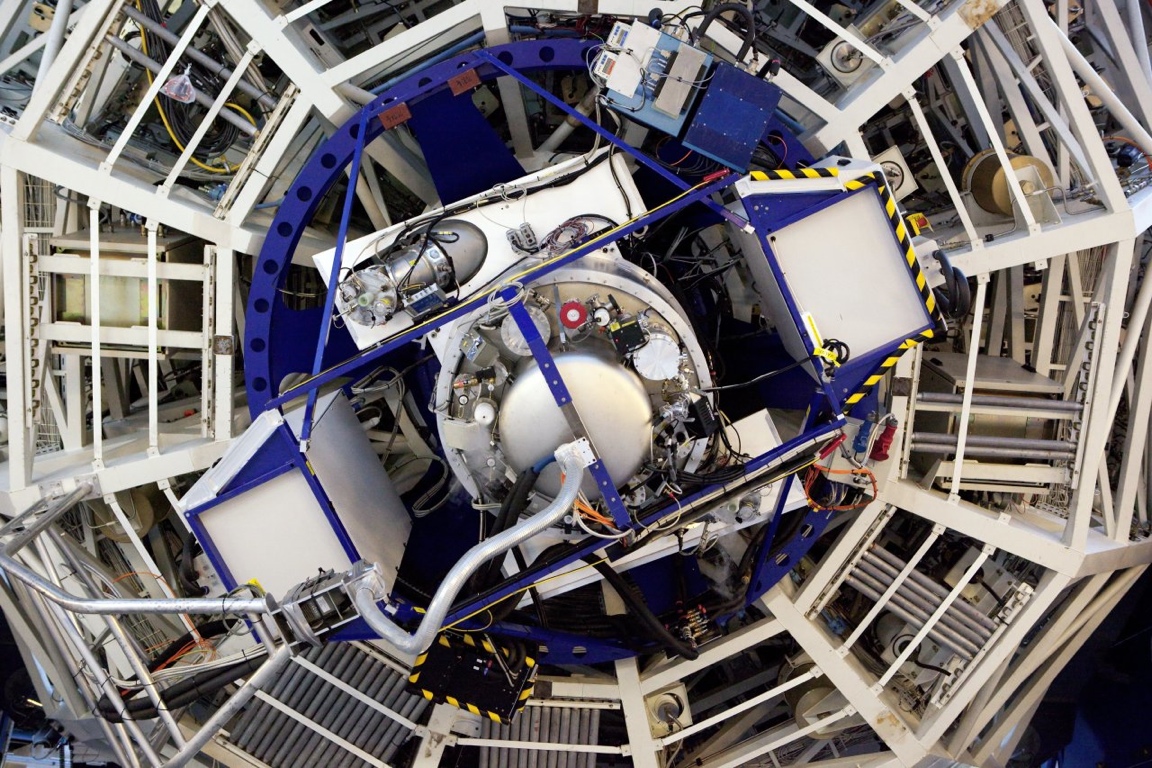}
\end{center}
 \caption{ \label{xsh_at_UT2} 
A view of X-shooter at the Cassegrain focus below the primary mirror
cell of the VLT UT2. In this view from below the instrument one sees the UVB
and VIS spectrographs at the top and bottom, respectively. The NIR
cryostat is visible in the center. The two boxes on the left and on the
right are electronic cabinets.}
 \end{figure}

 An overview of the instrument design is given in
 Sect.~\ref{description}.  Performance (resolution, throughput,
 background, stability) as measured during testing, commissioning and
 the first months of science operations is discussed in
 Sect.~\ref{performances}. In Sect.~\ref{conclusion} we summarize the
 conclusions and provide some suggestions for improvement of the
 instrument performance. We conclude in Sect.~\ref{example} with an
 example of a quasar observation obtained during commissioning as an
 illustration of the unique capabilities of the instrument.

\section{Instrument design}\label{description}

\begin{figure*}
\begin{center}
\includegraphics[width=14cm]{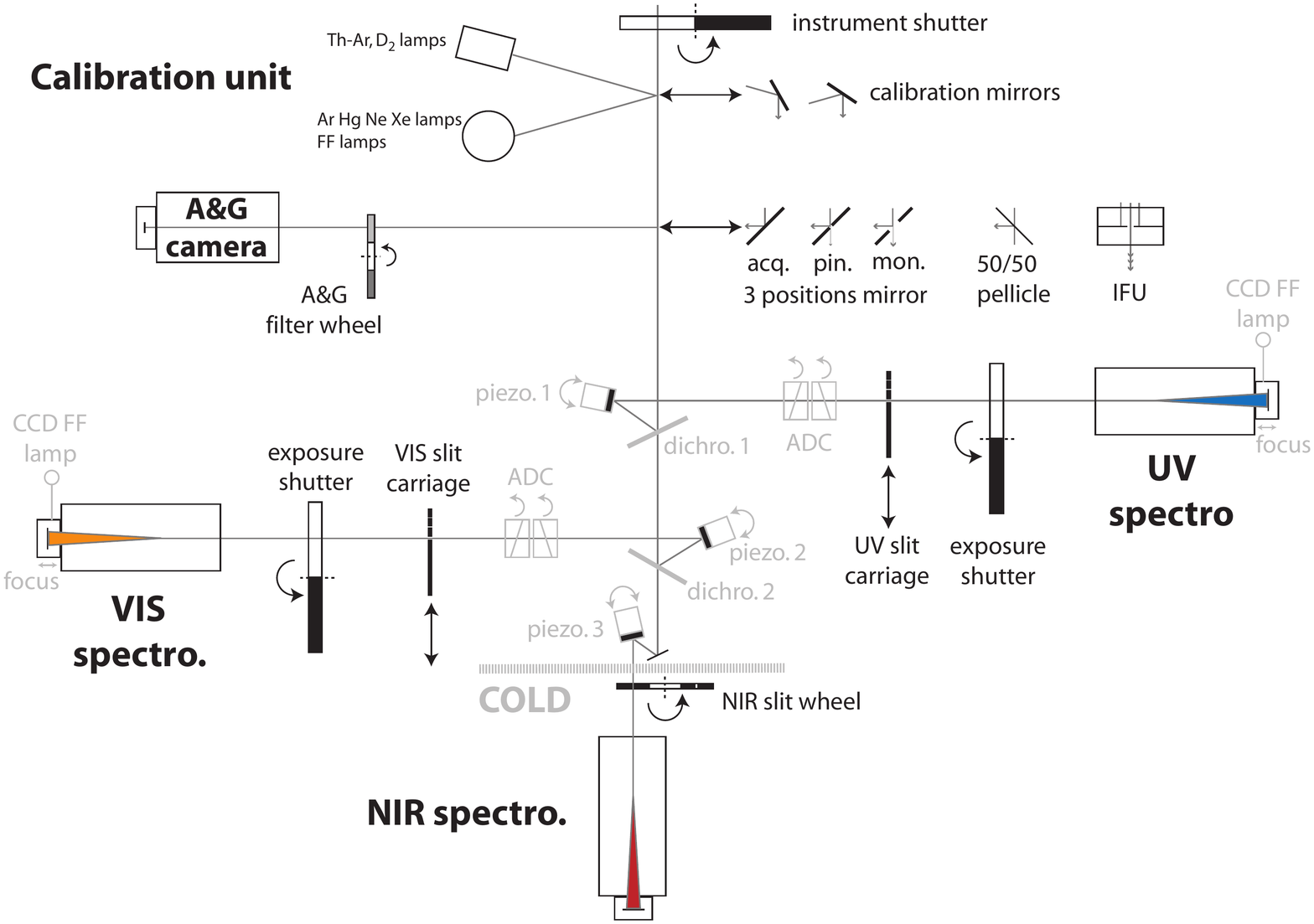}
\end{center}
\caption{ \label{schematic} Functional diagram of X-shooter. The light
  path runs from the top to the bottom of the figure. Each element is
  described in Sect.~\ref{description}.}
\end{figure*}

X-shooter consists of a central structure (the backbone) that
supports three prism-cross-dispersed \'echelle spectrographs optimized
for the UV-Blue (UVB), Visible (VIS) and Near-IR (NIR) wavelength
ranges, respectively. After the telescope focus, a series of two
highly efficient dichroics reflect the UVB and VIS light to the
corresponding arms and transmit longer wavelengths to the NIR arm. A
slicer can be inserted in the focal plane, which reformats
1.8\arcsec$\times$4\arcsec~on the sky into a
0.6\arcsec$\times$12\arcsec~long slit. A slit unit equipped with
11\arcsec~long slits of different widths is located at the entrance of
each spectrograph. A functional diagram of the
instrument is given in Fig.~\ref{schematic}.

In this section, we give an overview of the design of X-shooter
following photons coming from the telescope. For more detailed
discussions of specific aspects and the manufacturing process please refer to the
following publications: \cite{Spano2006} for the optical design;
\cite{Rasmussen2008} for the backbone and the UVB and VIS
spectrographs; \cite{Navarro2006, Navarro2008} for the NIR
spectrograph; \cite{roelfsema2008} for the cryogenic design;
\cite{Guinouard2006} for the Integral Field Unit; \cite{Vidali2006}
for the control software; \cite{Goldoni2006} and \cite{modigliani2010}
for the data reduction software.

\subsection{Key design choices}

A number of key design choices were made in the phases of the
project definition. Possibly the most crucial design choice was on the
method used to split the incoming beam from the telescope between the three
spectral arms. The option to use a single slit in the telescope focal
plane was rejected because of the difficulty of designing a highly
efficient relay system and atmospheric dispersion correction for the
full spectral range, and the need for work-packages with clean
interfaces to be handled by the different consortium partners, which
is not possible when spectrographs are sharing a single slit.  The
solution that was finally adopted is based on the sequential use of
two dichroics after the focal plane, used at 15\degr~rather than
45\degr~to minimize polarization effects. The beams toward the UVB and
VIS spectrographs are then deviated to 90\degr~with folding
mirrors. These two folding mirrors together with one in the NIR path
are actively controlled to compensate for small motions due to
flexures in the backbone of the instrument and guarantee that the
three target images all remain centered on the three slit units
as the telescope is tracking (see Sect.~\ref{bb_flex}).

The optical design allows the introduction of two short-wavelength
atmospheric dispersion correctors (ADC) and the focusing of the target on the
slit units at the entrance of the respective arms.

The size, weight and flexure restrictions implied a very compact
optical design of the spectrographs, requiring an efficient folding of
the light path, especially for the NIR-arm. The solution was found in
selecting the ``4C'' design described in \cite{delabre89}.

The inclusion of the K band was the subject of a complex
trade-off. With its uncooled optics in the pre-slit area the
instrument could not be optimized for a low thermal background. On the
other hand the K band did fit well in the spectral format on the
detector and had a potentially high efficiency. It was finally
decided to include the band, but its inclusion should not reduce the
performance in the J- and H-bands. It was also decided not to cool
the instrument pre-slit optics.

Another key design choice was the spectral resolution in the three
arms. The goal was to build an instrument which reaches the dark sky
noise limit in about 30 minutes, while still providing medium
resolution to do quantitative work on emission and absorption
lines. In the NIR the resolution of 5600 for 0.9\arcsec~slit
permits the full separation (and subtraction) of the sky emission lines.
At UVB and VIS wavelengths, specific scientific programs did call for
higher resolving power, e.g. to optimally measure abundances. The
final choices 
(see Table~\ref{resol}) 
are obviously a compromise to cover a broad range of astrophysical
programs.

\subsection{The backbone}

\subsubsection{The instrument shutter and the calibration unit}

In the converging beam coming from the telescope, the first element is
the telescope entrance shutter which allows safe daytime use of
X-shooter for tests and calibration without stray-light entering the
system from the telescope side.

This is followed by the calibration unit that allows selection from a set of
flat-fielding and wavelength calibration lamps carefully chosen to
cover the whole wavelength range \citep{saitta2008,kerber2008}. This unit consists of a mechanical structure
holding calibration lamps, an integrating sphere, relay optics that
simulate the f/13.6 telescope beam, and a mirror slide with 3
positions that can be inserted in the telescope beam:

\begin{itemize}
\item[$\bullet$] one free position for a direct feed from the telescope,
\item[$\bullet$]  one mirror that reflects the light from the
  integrating sphere equipped with:
\begin{itemize}
\item wavelength calibration Ar, Hg, Ne and Xe Penray lamps operating
simultaneously;
\item three flat-field halogen lamps equipped with different balancing filters to
optimize the spectral energy distribution for each arm.
\end{itemize}
\item[$\bullet$] one mirror which reflects light from:
\begin{itemize}
\item a wavelength calibration hollow cathode Th-Ar lamp;
\item a D$_2$ lamp for flat-fielding the bluest part of the UVB spectral range.
\end{itemize}
\end{itemize}

\subsubsection{The acquisition and guiding slide}

Light coming either directly from the telescope or from the
calibration unit described above arrives at the acquisition and
guiding slide (hereafter A\&G slide). This structure allows the insertion into the beam of one of the following components:
\begin{itemize}
\item[$\bullet$] a flat 45\degr~ mirror with three positions for:
\begin{itemize}
\item acquisition and imaging (labeled {\sc acq.} in
  Fig.~\ref{schematic}): the full
  1.5\arcmin$\times$1.5\arcmin~field of view is sent to the A\&G camera. This
  is the position used during all acquisition sequences;
\item spectroscopic observations and monitoring (labeled {\sc mon.} in
    Fig. \ref{schematic}): a slot lets the central 10\arcsec$\times$15\arcsec of
the field go through to the spectrographs while reflecting the peripheral field to
the A\&G camera. This is the position used for all science observations.
\item optical alignment and engineering purposes: a 0.5\arcsec pinhole
  producing an artificial star (labeled {\sc pin.} in
  Fig.~\ref{schematic}) is placed in the focal plane.
\end{itemize}
\item[$\bullet$] the Integral Field Unit (IFU, see Sect.~\ref{ifu});
\item[$\bullet$] a 50/50 pellicle beam splitter at 45\degr~ used to look down into the instrument with the A\&G
camera for engineering purposes.
\end{itemize}

\subsubsection{The IFU}\label{ifu}

\begin{figure}
\begin{center}
\includegraphics[width=8.3cm]{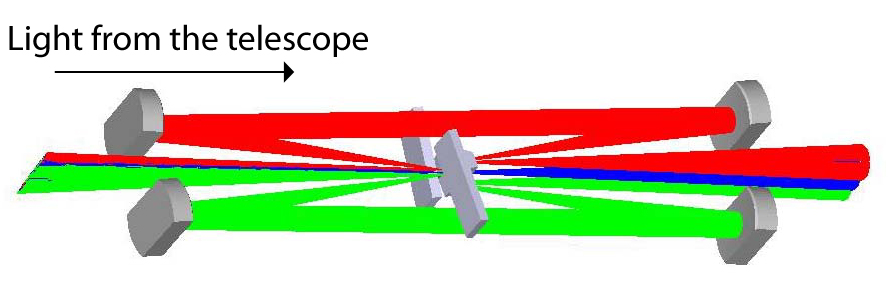}
\includegraphics[height=4.2cm]{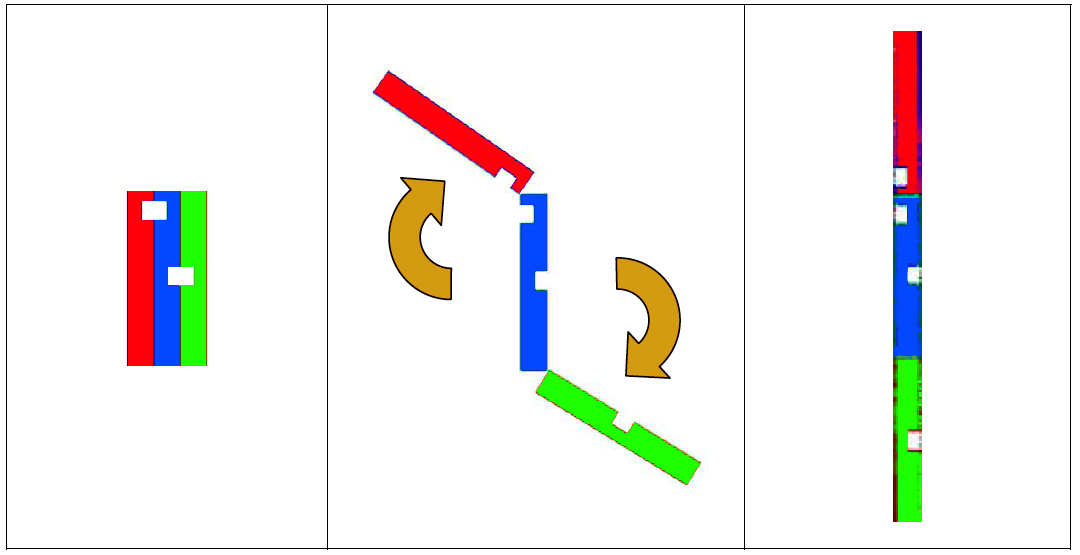}
\end{center}
\caption{ \label{ifu_fig} \textit{Top}: view of the effect of the
  IFU. 
The central field
  is directly transmitted to form the central slitlet (blue) while
  each lateral field (in red and green) is reflected toward a pair of
  spherical mirrors, and realigned at the end of the central slice to
  form the exit slit. \textit{Bottom}: The field before (left) and
  after the IFU (right). The IFU acts such that the lateral fields are
  rotated. The two white slots are not real gaps but just guides to
  help visualize the top and the bottom of each slice in the drawing.}
\end{figure}

The Integral Field Unit is an image slicer that re-images an input
field of 4\arcsec$\times$1.8\arcsec~into a pseudo slit of
12\arcsec$\times$0.6\arcsec. The light from the central slice is
directly transmitted to the spectrographs. The two lateral sliced
fields are reflected toward the two pairs of spherical mirrors and
re-aligned at both ends of the central slice in order to form the exit
slit as illustrated in Fig. \ref{ifu_fig}. Due to these four
reflections the throughput of the two lateral fields is reduced with
respect to the directly transmitted central one. The measured overall
efficiency of the two lateral slitlets is ~85\% of the direct
transmission but drops to ~50\% below 400 nm due to reduced coating
efficiency in the blue. Note that each spectrograph is equipped with a
dedicated 12.6\arcsec$\times$1\arcsec\, opening to be used in
combination with the IFU. It is slightly larger than
the IFU pseudo slit ensuring that the whole field of view is
transmitted while baffling ghosts.

\subsubsection{The acquisition and guiding camera}

The A\&G camera allows visual detection and centerering of objects from the U- to the z-band.
This unit consists of:
\begin{itemize}
\item[$\bullet$] a filter wheel equipped with a full UBVRI Johnson filter set and a full Sloan Digital
Sky Survey (SDDS) filter set.
\item[$\bullet$] a Peltier cooled, 13~$\mu$m pixel, 512$\times$512 E2V
  broad band coated Technical CCD57\,--\,10 onto which the focal plane
  is re-imaged at f/1.91 through a focal reducer. This setup provides
  a plate scale of 0.173\arcsec/pix and a field of view of
  1.47\arcmin$\times$1.47\arcmin.
\end{itemize}
This acquisition device ---that can also be used to record images of the target field through
different filters--- provides a good enough sampling to measure the
centroid of a target to better than 0.1\arcsec~accuracy in all
seeing conditions.

\subsubsection{The dichroic box}

\begin{figure}
\begin{center}
\includegraphics[width=9cm,trim =5mm 5mm 5mm 5mm, clip]{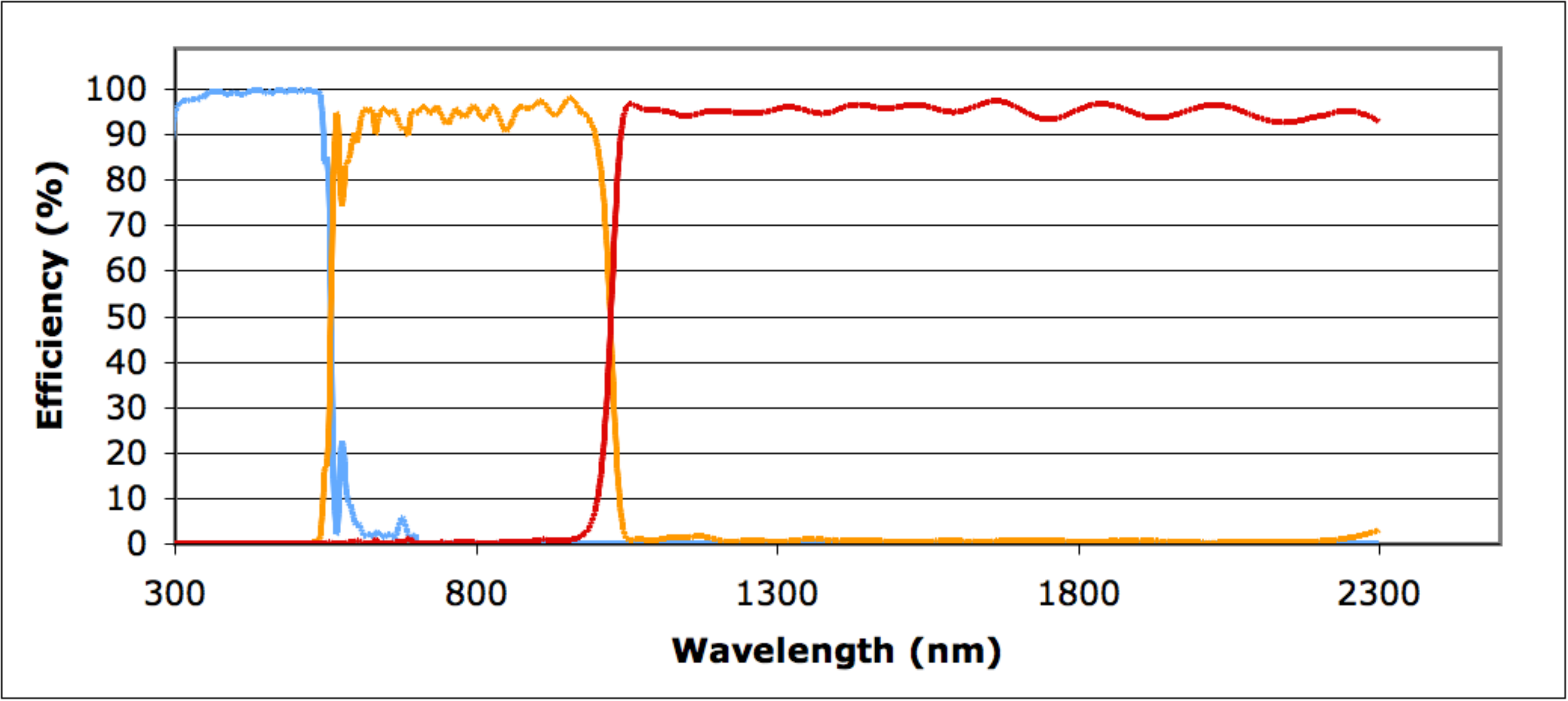}
\end{center}
\caption{ \label{dic_fig} 
The combined efficiency of the two dichroic beam splitters. In blue: reflection
on the first dichroic; in orange: transmission through the
first dichroic and reflection on the second dichroic; in
red: transmission through both dichroics.}
\end{figure} 

Light is split and distributed to the three arms by two highly
efficient dichroic beam splitters.  These are the first optical
elements encountered by the science light (unless the IFU is
deployed). The first dichroic at an incidence angle of 15\degr~
reflects more than 98\% of the light between 350 and 543 nm and
transmits $\sim$95\% of the light between 600 and 2300 nm. The second
dichroic, also at 15\degr~ incidence, has a reflectivity above 98\%
between 535 nm and 985 nm and transmits more than 96\% of the light
between 1045 and 2300 nm. The combined efficiency of the two dichroics
is shown in Fig.~\ref{dic_fig}: it is well above 90\% over most of the
spectral range.

\subsubsection{The flexure compensation tip-tilt mirrors}\label{tiptilt}

Light reflected and/or transmitted by the two dichroics reaches, in
each arm, a folding mirror mounted on a piezo tip-tilt mount (S-340
from Physik Instrumente). These mirrors are used to fold the beam and
correct for backbone flexure to keep the relative alignment of the
three spectrograph slits fixed at any orientation of the telescope and
instrument. Operational aspects and performance of the flexure
compensation system are addressed in Sect.~\ref{bb_flex}.

For slit observations (but not IFU) these tip-tilt mirrors also
compensate for shifts due to atmospheric differential refraction
between the telescope tracking wavelength (fixed at 470 nm) and the
undeviated wavelength of the two Atmospheric Dispersion Correctors
(for UVB and VIS arms, see Sect.~\ref{adc}) and the middle of the atmospheric dispersion
range for the NIR arm.

\subsubsection{\label{adc}The focal reducer and atmospheric dispersion
  correctors}

Both UVB and VIS pre-slit arms contain a focal reducer and an
atmospheric dispersion corrector (ADC). These focal reducer-ADCs
consist of two doublets cemented onto two counter rotating double
prisms. The focal reducers bring the focal ratio from f/13.41 to
~f/6.5 and provide a measured plate scale at the entrance slit of the
spectrographs of 3.91\arcsec/mm in the UVB and 3.82\arcsec/mm in the
VIS. The ADCs compensate for atmospheric dispersion in order to
minimize slit losses and allow orienting the slit to any position
angle on the sky up to a zenith distance of 60\degr. The zero
deviation wavelengths are 405 and 633 nm for the UVB and the VIS ADCs,
respectively. During slit observations, their positions are updated
every 60s based on information taken from the telescope
database.

Since the IFU comes ahead of the ADCs in the optical train, no
correction for atmospheric dispersion is available for IFU
observations, and the ADCs are set to their neutral position in this
observing mode.

The NIR arm is not equipped with an ADC. The NIR arm tip-tilt mirror
compensates for atmospheric refraction between the telescope tracking
wavelength (470 nm) and 1310 nm which corresponds to the middle of the
atmospheric dispersion range for the NIR arm. This means that this
wavelength is kept at the center of the NIR slit. At a zenith
distance of 60\degr~the length of the spectrum dispersed by the
atmosphere is 0.35\arcsec, so the extremes of the spectrum can be
displaced with respect to the center of the slit by up to
0.175\arcsec.

\subsection{The UVB spectrograph}

\subsubsection{Slit carriage}\label{uvb_slit}
The first opto-mechanical element of the spectrograph is the slit
carriage. Besides the slit selection mechanism, this unit consists of
a field lens placed just in front of the slit to re-image the telescope
pupil onto the spectrograph grating, and the spectrograph shutter just
after the slit. The slit mask is a laser cut Invar plate manufactured
with a LPKF Laser Cutter. It is mounted on a
motorized slide in order to select one of the 9 positions
available. All science observation slits are 11\arcsec\,high and different
widths are available: 0.5\arcsec, 0.8\arcsec, 1.0\arcsec, 1.3\arcsec,
1.6\arcsec\,and 5\arcsec\,(the latter for spectro-photometric
calibration, see Table~\ref{resol}). In addition a single pinhole for spectral format check
and order tracing and a 9-pinhole mask for wavelength calibration and
spatial scale mapping are available. A 12.6\arcsec$\times$1\arcsec\,
slit is also available to be used in combination with the IFU (see Sect.~\ref{ifu}).

\subsubsection{Optical layout}

\begin{figure}
\begin{center}
\includegraphics[width=9cm]{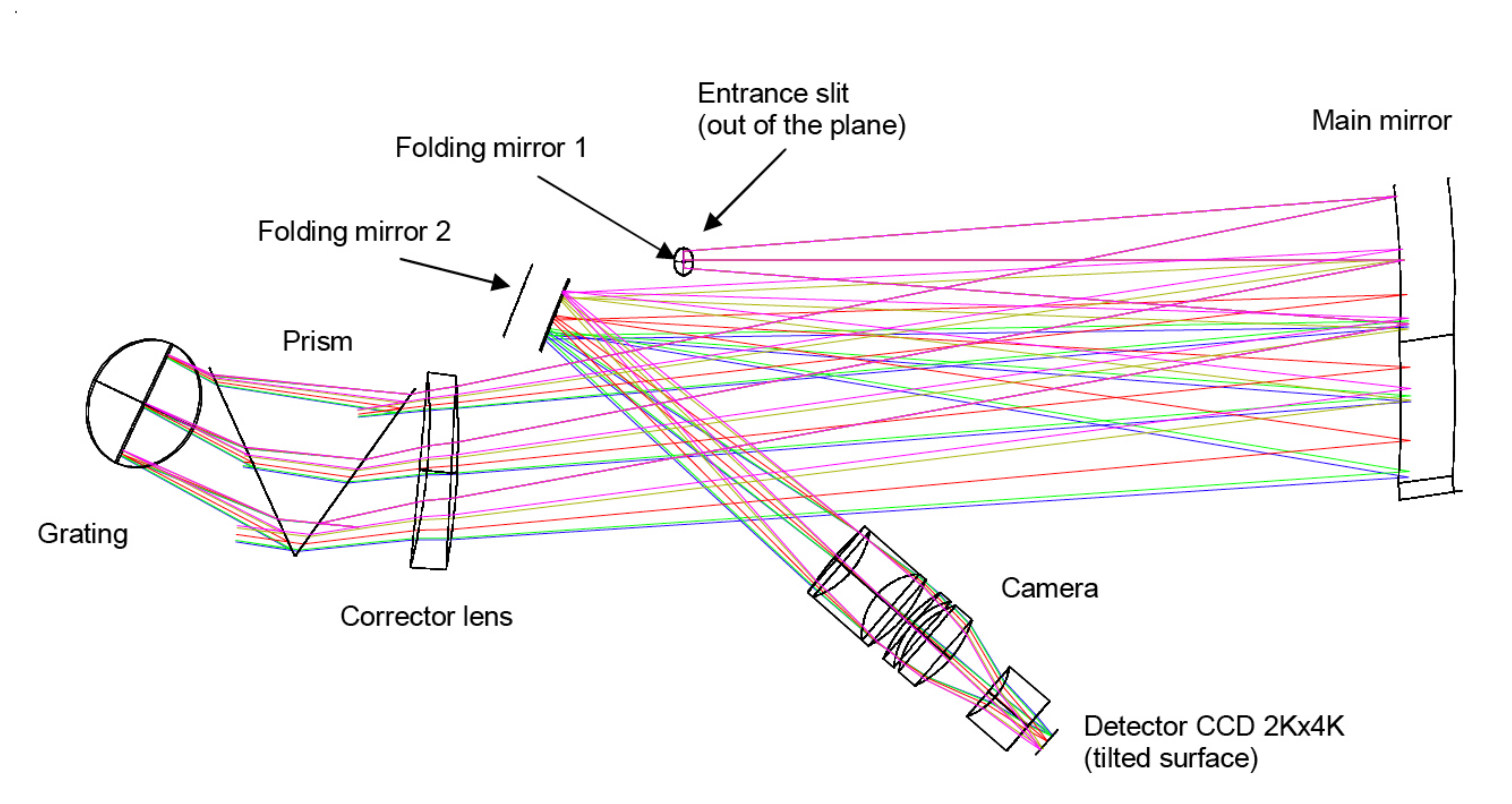}
\end{center}
\caption{ \label{uvb_opt} 
The UVB spectrograph optical layout. The optical layout of the VIS
spectrograph is very similar to this one.
}
\end{figure} 

The optical layout of the UVB spectrograph is presented in
Fig.~\ref{uvb_opt}. Light from the entrance slit, placed behind the
plane of the figure, feeds a 5\degr\,off-axis Maksutov-type collimator
through a folding mirror. The collimator consists of a spherical
mirror and a diverging fused silica (SiO$_2$) corrector lens with only spherical
surfaces. The collimated beam passes through a 60\degr~silica prism twice
to gain enough cross-dispersion. The main dispersion is achieved through a
180 grooves/mm \'echelle grating blazed at 41.77\degr. The off-blaze
angle is 0.0\degr, while the off-plane angle is 2.2\degr. After
dispersion, the collimator creates an intermediate spectrum near the
entrance slit, where a second folding mirror has been placed. This
folding mirror acts also as a field mirror. Then a dioptric camera (4
lens groups with CaF$_{2}$ or silica lenses, one aspherical surface) re-images
the cross-dispersed spectrum at f/2.7 (plate scale 9.31\arcsec/mm)
onto a detector that is slightly tilted to compensate for a variation
of best focus with wavelength. The back focal length is rather
sensitive to temperature changes. It varies by $\sim$22.7~$\mu$m/\degr C which
corresponds to a defocus of 9~$\mu$m/\degr C or
$\sim$0.08\arcsec/\degr C. This is automatically compensated for at the
beginning of every exposure by moving the triplet+doublet of the
camera by $-$10.9~$\mu$m/\degr C.

\subsubsection{Detector}

The UVB detector is a 2048$\times$4102, 15~$\mu$m pixel CCD from E2V
(type CCD44-82) of which only a 1800$\times$3000 pixels window is
used. The CCD cryostat is attached to the camera with the last optical
element acting as a window. The operating temperature is 153~K. The CCD
control system is a standard ESO FIERA controller
\citep[see][]{beletic98} shared with the VIS CCD. The associated
shutter, located just after the slit, is a 25~mm bi-stable shutter from
Uniblitz (type BDS 25). Full transit time is 13~ms. Since the slit is
2.8~mm high (11\arcsec at f/6.5), the illumination of the detector is
homogeneous to within $\ll$10~ms.

\subsection{The VIS spectrograph}
\subsubsection{Slit carriage}

The slit carriage of the VIS spectrograph is identical to that of the
UVB arm (see Sect.~\ref{uvb_slit}), but the available slits are
different. All the science observation slits are 11\arcsec~high and
the slit widths are: 0.4\arcsec, 0.7\arcsec, 0.9\arcsec,
1.2\arcsec, 1.5\arcsec~and 5\arcsec (see
Table~\ref{resol}). 

\subsubsection{Optical layout}
The optical layout of the VIS spectrograph is very similar to that of
the UVB (see Fig.~\ref{uvb_opt}).  The collimator (mirror+corrector
lens) is identical. For cross-dispersion, it uses a 49\degr~ Schott
SF6 prism in double pass. The main dispersion is achieved through a
99.4 grooves/mm, 54.0\degr~ blaze \'{e}chelle grating. The off-blaze angle
is 0.0\degr~ and the off-plane angle is 2.0\degr. The camera (three lens
groups, one aspherical surface) re-images the cross-dispersed spectrum at
f/2.8 (plate scale 8.98\arcsec/mm) onto the detector (not
tilted). Focussing is obtained by acting on the triplet+doublet
sub-unit of the camera. However, unlike the UVB arm, the back focal
length varies by less than 1~$\mu$m/\degr C (image blur $<$0.004\arcsec/\degr
C) hence no thermal focus compensation is needed.

\subsubsection{Detector}
The VIS detector is a 2048$\times$4096, 15~$\mu$m pixel CCD from MIT/LL
(type CCID-20). As in the UVB arm, the cryostat is attached to the
camera with the last optical element acting as a window. The operating
temperature is 135~K. It shares its controller with the UVB
detector. The associated shutter system is identical to the UVB one.

\subsection{The NIR spectrograph}

The NIR spectrograph is fully cryogenic. It is cooled with a liquid
nitrogen bath cryostat and operates at 105~K. 

\subsubsection{Pre-slit optics and entrance window}
After the dichroic box and two warm mirrors M1 (cylindrical) and M2
(spherical, mounted on a tip-tilt stage and used for flexure
compensation (see Sect.~\ref{tiptilt}), light enters the cryostat via the
Infrasil vacuum window. To avoid ghosts, this window is tilted by
3\degr. After the window, light passes the cold stop, and is directed
towards the slit via two folding mirrors M3 (flat) and M4
(spherical).

\subsubsection{Slit wheel}
A circular laser cut Invar slit mask is pressed in between two
stainless steel disks with 12 openings forming the wheel. The wheel is
positioned by indents on the circumference of the wheel with a roll
clicking into the indents. All the science observation slits are
11\arcsec~high and different widths are offered: 0.4\arcsec,
0.6\arcsec, 0.9\arcsec, 1.2\arcsec, 1.5\arcsec~and 5\arcsec (see
Table~\ref{resol}). As for the two other arms, a single pinhole, a 9
pinhole mask and an IFU dedicated 12.6\arcsec$\times$1\arcsec slit are available.

\subsubsection{Optical layout}

\begin{figure}
\begin{center}
\includegraphics[width=9cm]{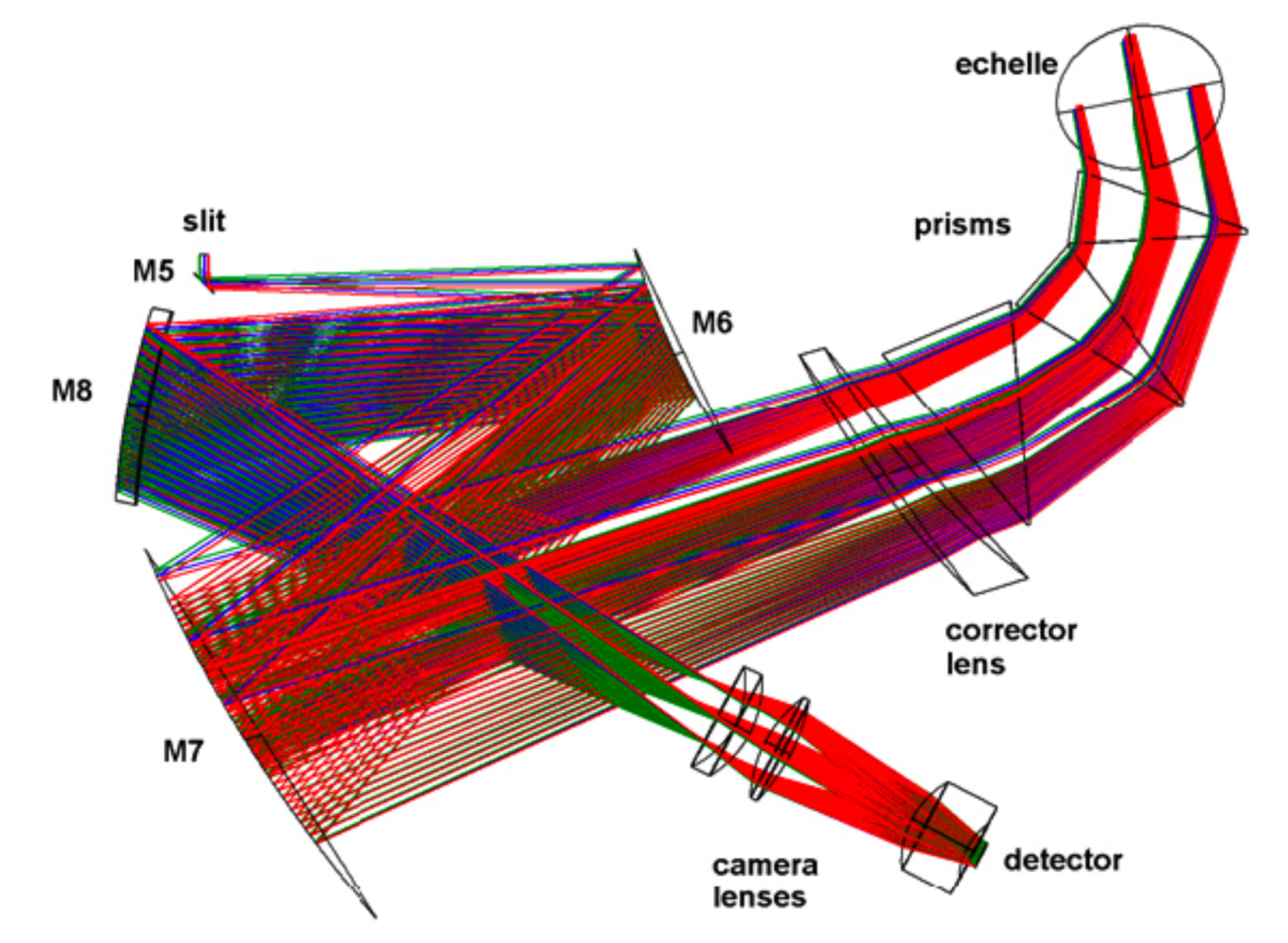}
\end{center}
\caption{ \label{nir_opt} 
The NIR spectrograph optical layout.
}
\end{figure} 

The optical layout of the NIR spectrograph is presented in
Fig.~\ref{nir_opt}. The conceptual design is the same as for the UVB
and the VIS spectrographs. Light entering the spectrograph via the
entrance slit and folding mirror M5 feeds an off-axis
Maksutov-inspired collimator. In this case, the collimator is made of
two spherical mirrors M6 and M7 plus an Infrasil corrector lens (with
only spherical surfaces). In order to get enough cross dispersion,
three prisms are used in double pass. Prism 1 is a 35\degr~top angle
made of Infrasil; prisms 2 and 3 are two 22\degr~top angle ZnSe prisms
of the largest available thickness (56~mm). This design
provides an almost constant order separation. The main dispersion is
provided by a 55 grooves/mm \'echelle grating with a blaze angle of
46.07\degr. The off-blaze angle is 0.0\degr, while the off-plane angle
is 1.8\degr. After dispersion, the collimator creates an intermediate
spectrum near the entrance slit, where M8, a spherical mirror, acts as
a field mirror, relocating the pupil between L2 and L3, the last
lenses of the camera. The fixed focus camera re-images the
\'{e}chellogram onto the detector at f/2.1 (plate scale
12.1\arcsec/mm).

The optical box and the mechanical structures of the
optical elements are made out of a single block of aluminum, designed
such that after cooling down the optical elements are accurately
positioned taking into account the shrinkage of the aluminum. To
reduce flexure, extreme lightening techniques have been applied
to reduce the weight of the NIR spectrograph at the bottom of the
instrument in the Cassegrain focus.

\subsubsection{\label{nir_det} Detector}

The NIR detector is a Teledyne substrate-removed HgCdTe, 2k$\times$2k,
18~$\mu$m pixel Hawaii 2RG of which only 1k$\times$2k is used. It
is operated at 81~K. Cooling is achieved through a heat link (a
massive 40~mm$\times$40~mm copper bar) plunged directly into the
bottom of the liquid N$_2$ tank. The array control system is the ESO
standard IRACE controller \citep{meyer98}.  Sample-up-the-ramp
(non-destructive) readout is always used. This means that during
integration, the detector is continuously read out without resetting
it and counts in each pixel are computed by fitting the slope of the
signal vs. time.  In addition, Threshold Limited Integration (TLI)
mode is used to extend the dynamical range for long exposure times: if

one pixel is illuminated by a bright source and reaches an absolute
value above a certain threshold (close to detector saturation), only
detector readouts before the threshold is reached are used to compute
the slope and the counts written in the FITS image for this pixel are
extrapolated to the entire exposure time \citep[see][]{finger2008}.

To significantly decrease persistence, a global reset is applied
immediately after finishing science exposures and pixels of the array
are kept at the reset voltage until the next exposure starts. The
release of trapped charge during a dark exposure immediately following
an exposure to a bright light source is the cause of the persistence
effect. If all pixels of the array are connected to the reset voltage
the diode junctions and the width of their depletion regions do not
change even if the array is exposed to a bright light source with
photons generating charge. Hence, no traps in the depletion region are
exposed to majority carriers. On the contrary, trapped charge is
released during the global reset before the next exposure starts.  By
two minutes of global reset de-trapping (typical time interval in
between two science exposures in a nodding sequence) the persistence
effect can be reduced by a factor of ten. If the reset switch is kept
closed during the bright exposure prior to the dark exposures the
persistence is eliminated and the global reset acts like an electronic
shutter protecting the detector from exposure to bright sources.

\subsection{The Instrument Control and Observing Software}

The X-Shooter control software is based on the standard ESO VLT
control software \citep[see][]{raffi97}. The number of functions to
control is relatively low compared to other large VLT instruments (8
calibration lamps; 13 motors; 3 tip-tilt piezoelectric actuators; 55
digital and analog sensors and 4 detectors, see
Fig.~\ref{schematic}). The two most critical aspects are: the
synchronization of exposures between the three arms and the real time
flexure compensation, see Sect.~\ref{bb_flex}.

\subsection{Data reduction software}
 
\begin{figure*}
\begin{center}
\includegraphics[width=6cm]{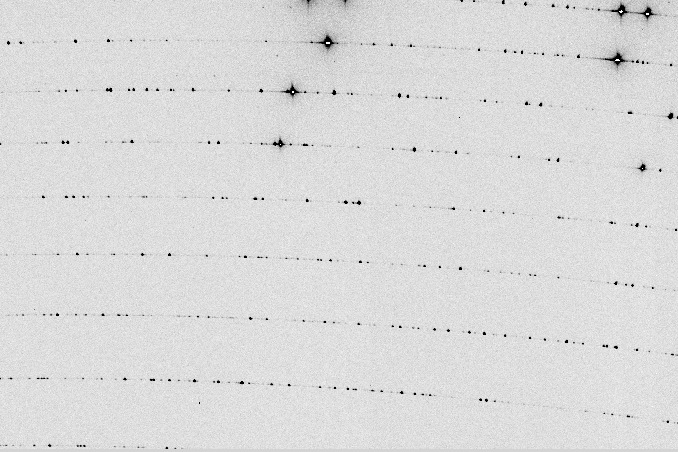}
\includegraphics[width=6cm]{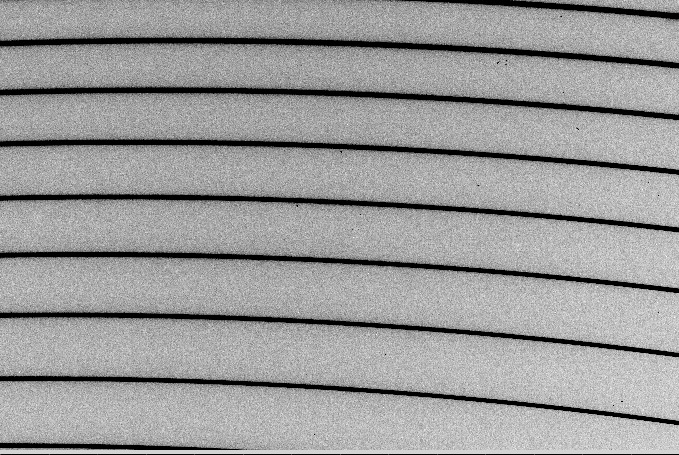}
\includegraphics[width=6cm]{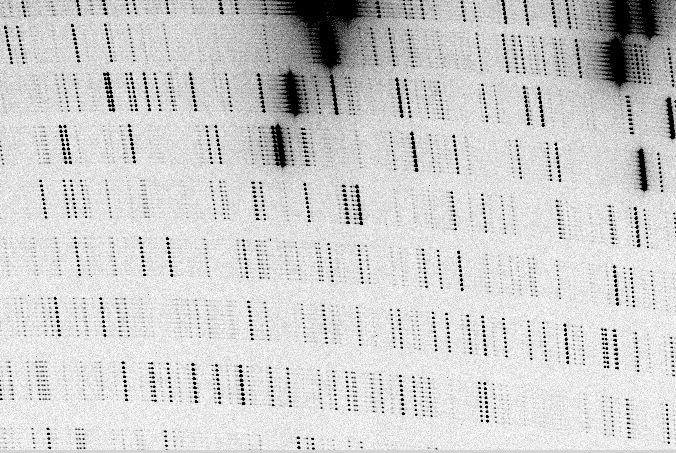}
\end{center}
\caption{ \label{calib} Sections 
(1365 pix $\times$ 925 pix)  of VIS arm calibration frames used by
  the data reduction pipeline to fully characterize the spectral
  format. From left to right: a single pinhole arc frame (``format
  check''), a single pinhole continuum frame ("order definition") and
  a 9-pinhole arc frame used for spatial scale and wavelength
  calibration.}
\end{figure*}

Delivery of the X-shooter data reduction
pipeline\footnote{available at
  http://www.eso.org/sci/software/pipelines/}
\citep{Goldoni2006, modigliani2010} was an important part of the
X-shooter project. It provides recipes for Paranal Science Operations,
for data Quality Control at ESO headquarters and for offline data
reduction by science users. While it is used in a fully automated mode
for quick-look data evaluation on Paranal, the pipeline is fully
configurable through an extended set of parameters to allow
astronomers to tune the reduction to their specific needs. Pipeline
recipes can be executed either with the ESO Recipe Execution Tool
(EsoRex\footnote{available at http://www.eso.org/sci/software/cpl/esorex.html}) at
the command line level or through the Gasgano\footnote{available at
  http://www.eso.org/sci/software/gasgano/} graphical user
interface. The recipes are implemented with the ESO Common Pipeline
Library \citep[CPL, ][]{banse2004,mckay2004}.

The data reduction pipeline is built up in modules which in
combination lead to fully reduced spectra but permit the extraction of
intermediate results when required by the user. Errors are propagated
throughout the whole reduction chain \citep[see][]{horrobin08}. The
most important modules are described below:

\begin{itemize}
\item {\bf xsh\_mbias} and {\bf xsh\_mdark} combine series of raw
  biases and darks into a master bias and a master dark
  respectively. These steps also update a reference master bad pixel
  map.
\item \textbf{xsh\_predict} takes as input a single pinhole arc
  calibration frame (named format check, see left hand panel on
  Fig~\ref{calib}) and computes a first guess for the wavelength
  solution and position of the center of each order taking into
  account information on atmospheric pressure and instrument
  temperature available in the FITS header.
\item \textbf{xsh\_orderpos} takes as input a continuum lamp
  illuminated single pinhole calibration frame  (named order
  definition frame, see central panel on Fig~\ref{calib}) and accurately traces
  the center of each order.  
 \item \textbf{xsh\_mflat} combines a series of raw flat field frames into a
   master flat field. It also traces the edge of each orders. In case
   of IFU calibrations, it traces the edge of each slitlet. 
 \item \textbf{xsh\_2dmap} takes as input a 9-pinhole mask arc frame
   (see an example on Fig~\ref{calib}, right hand panel) together with a first
   guess wavelength solution derived by the preceding recipes and determines
   the wavelength and spatial scale calibration. Two calibration
   methods are proposed: either a classical method based on two
   dimensional polynomial fitting of the detected arc lines or a method
   based on the optimization of a physical model of the instrument
   \citep[see][]{bristow2008}.
 \item \textbf{xsh\_flexcomp} updates the wavelength solution to
   correct for the effect of flexures and temperature drifts using the
   first frame of the active flexure compensation sequence taken
   during each pointing (see Sect.~\ref{bb_flex}).
\item \textbf{xsh\_response} takes as input observations of a
  spectro-photometric standard star and computes the response function.
\item \textbf{xsh\_scired\_slit\_stare, xsh\_scired\_slit\_nod} and
  \textbf{xsh\_scired\_slit\_onoff} recipes process science data for
  the three main observing strategies used in slit mode: staring,
  nodding along the slit and sampling of the sky off target,
  respectively. These recipes first subtract the bias (UVB and VIS) or the
  master dark (NIR) and divide by a master flat-field. When less than
  three science frames per pointing are given as input, cosmic ray
  rejection using laplacien edge detection method described in
  \cite{vandokkum01b} is applied to each frame, otherwise frames are
  combined with a Kappa-Sigma clipping. In the staring case, the sky
  background is fitted and subtracted taking advantage of the fine
  sampling naturally provided by distortions of the spectral format
  following prescriptions detailed in \cite{kelson2003}. In case of
  nodding along the slit, the so-called double pass sky subtraction is
  applied (first pass: subtraction of frames from the second position
  from those of the first one to produce a difference frame; second
  pass: co-addition of the difference frame with a negative and
  shifted version of itself to co-add signal from the two positions
  and remove residuals due to variations in the sky background). These
  recipes produce, for each arm, a two-dimensional rectified spectrum that is
  wavelength, spatial scale and flux calibrated. It is possible to
  automatically detect objects and extract one-dimensional spectra
  with either a simple sum over an aperture or an optimal extraction.
\item \textbf{xsh\_scired\_ifu\_stare} and
  \textbf{xsh\_scired\_ifu\_onoff} process IFU data and reconstruct
  calibrated data cubes for each arm. These recipes are similar to
  their slit mode equivalents described above. 
The flat-fielding step corrects
  for the difference in
  throughput between the two lateral sub-fields and the central one
  (see Section~\ref{ifu}).
 Note that due to the
  small field of view of the IFU, nodding within the IFU is not a
  recommended observing strategy and thus is not supported by the
  pipeline.
\end{itemize}

\section{X-shooter performance}\label{performances}

\subsection{Detectors}

The performance of the detectors of each arm are given in Table~\ref{det_perf} for
all readout modes offered for science observations and calibrations.

\begin{table*}
  \caption{\label{det_perf}Detector performances.}
  \centering
  \begin{tabular}{c c c c}
\hline
\hline
   Chanel & UVB  & VIS & NIR \\
\hline
   Detector type & e2v CCD44-82 & MIT/LL CCID 20 & Hawaii 2RG\\
&&&(substrate removed)\\
\hline
  QE & 80\% at 320 nm  & 78\% at 550nm & 85\% \\
       &88\% at 400 nm & 91\% at 700nm &\\
       &83\% at 500 nm & 74\% at 900nm &\\
       &81\% at 540 nm & 23\% at 1000nm &\\
\hline
Gain & High: 0.62 & High 0.595 & 2.12\\
(e$^{-}$/ADU) & Low: 1.75 & Low: 1.4 &\\
\hline
Readout noise & Slow: 2.5 & Slow 3.1 & Short DIT: $\sim$25\\
(e$^{-}$ rms) & Fast: 4.5 & Fast: 5.2 &DIT$>$300s:$\sim$8\\
\hline
Full frame     &1$\times$1, slow-fast: 70-19&1$\times$1, slow-fast: 92-24&0.665 (for a\\
readout time &1$\times$2, slow-fast: 38-12&1$\times$2, slow-fast: 48-14&single readout)\\
(s)                 &2$\times$2, slow-fast: 22-8  &2$\times$2, slow-fast: 27-9&\\
\hline
Dark current &$<$\,0.2&$<$1.1&21\\ 
(e$^{-}$/pix/h)&&&\\
\hline
Fringing &&$\sim$5\% peak-to-valley&\\
amplitude&&&\\
\hline
Non-linearity&Slow: 0.4\%&Slow:0.8\%&$<$1\% up to \\
                     &Fast: 1.0\%&Fast: 0.8\%&45000 ADU\\
\hline
  \end{tabular} 
\end{table*}

\subsection{Spectral format}

The spectral format of X-shooter is fixed. The whole spectral range is
covered by 12 orders in the UVB, 15 in the VIS, and 16 in the
NIR. Orders are strongly curved (parabolic) and the spectral line tilt
varies along the orders. Both slit height and width projection also vary
from order to order and along each order due to a variable anamorphic
effect introduced by the prisms (crossed twice). The minimum
separation between orders is $\sim$4 (unbinned) pixels to allow inter-order
background evaluation. The dichroic crossover between UVB-VIS and
VIS-NIR is at 559.5 nm and 1024 nm respectively, near the location of
atmospheric features. Grating line densities were chosen to have the
crossovers occur near the ends of the order. The spectral ranges on
the detector and blaze wavelength for each order are given in
Table~\ref{spec_fmt} together with an example of a Th-Ar slit frame for
each arm. These measurements are in excellent agreement with the
predicted spectral format \citep{Spano2006}.

\begin{table*}
\caption{\label{spec_fmt} {The X-shooter spectral format for the UVB
(top), VIS (middle) and NIR (bottom) arm as measured at the
telescope. The minimum and maximum wavelength recorded on the detector
together with the blaze wavelength are given for each order ; on the right column, 
an example of wavelength calibration frame taken with a ThAr lamp for
each arm.}}
\centering
\begin{tabular}{c c c c c}
\hline\hline
    Order& min. wavelength & Blaze wavelength & Max. wavelength & Example of a  ThAr       \\
             &(nm)                    &(nm)                      &(nm)                     &       calibration frame                               \\
\hline
\multicolumn{5}{c}{\textbf{UVB}}\\
\hline          
24&	293.6&	312.2&	322.3&\multirow{12}{*}{\includegraphics[height=4cm]{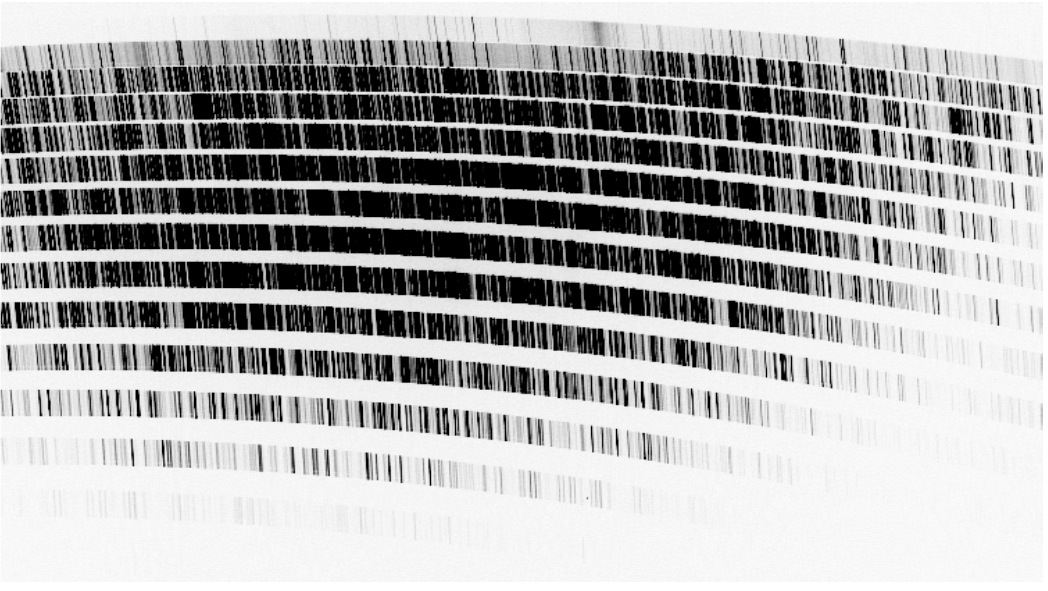}}\\
23&	306.2&	325.0&	336.2&\\
22&	320.0&	339.8&	351.4&\\
21&	335.1&	356.1&	368.0&\\
20&	351.8&	373.5&	386.2&\\
19&	370.1&	393.2&	406.4&\\
18&	390.6&	414.5&	428.9&\\
17&	413.4&	438.8&	454.0&\\
16&	439.1&	466.4&	482.2&\\
15&	468.3&	496.8&	514.2&\\
14&	501.6&	531.0&	550.8&\\
13&	540.1&	556.0&	593.0&\\
\hline
\multicolumn{5}{c}{\textbf{VIS}}\\
\hline          
30&	525.3&	550.5&	561.0&\multirow{16}{*}{\includegraphics[height=4.cm]{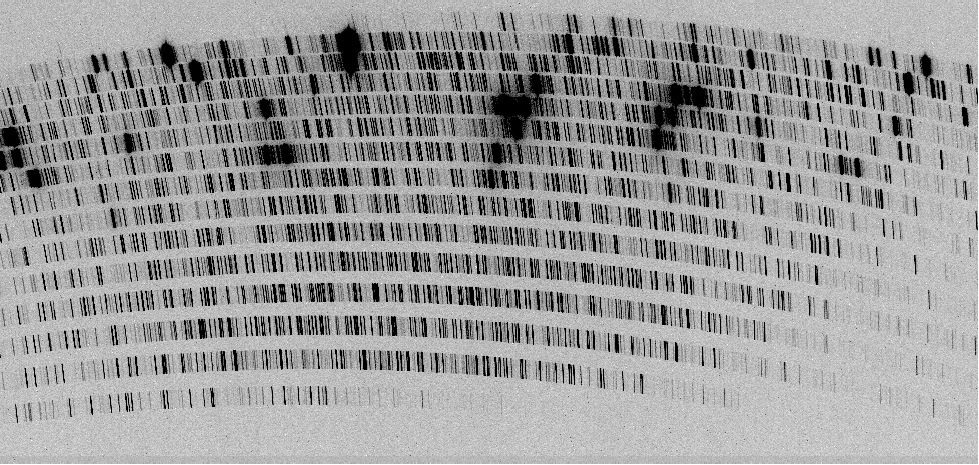}}\\
29&	535.8&	568.0&	580.2&\\
28&	554.6&	585.9&	600.8&\\
27&	575.2&	607.7&	622.9&\\
26&	597.4&	629.5&	646.8&\\
25&	621.3&	653.8&	672.5&\\
24&	647.2&	682.1&	700.4&\\
23&	675.4&	711.2&	730.7&\\
22&	706.1&	742.6&	763.8&\\
21&	739.7&	777.6&	800.0&\\
20&	777.0&	815.8&	839.8&\\
19&	817.6&	860.2&	883.8&\\
18&	862.9&	904.3&	932.7&\\
17&	913.7&	957.3&	987.4&\\
16&	970.7&	1001.6&	1048.9&\\
\hline
\multicolumn{5}{c}{\textbf{NIR}}\\
\hline          
26&	982.7	&1005.8	&1034.2&\multirow{16}{*}{\includegraphics[height=4.2cm]{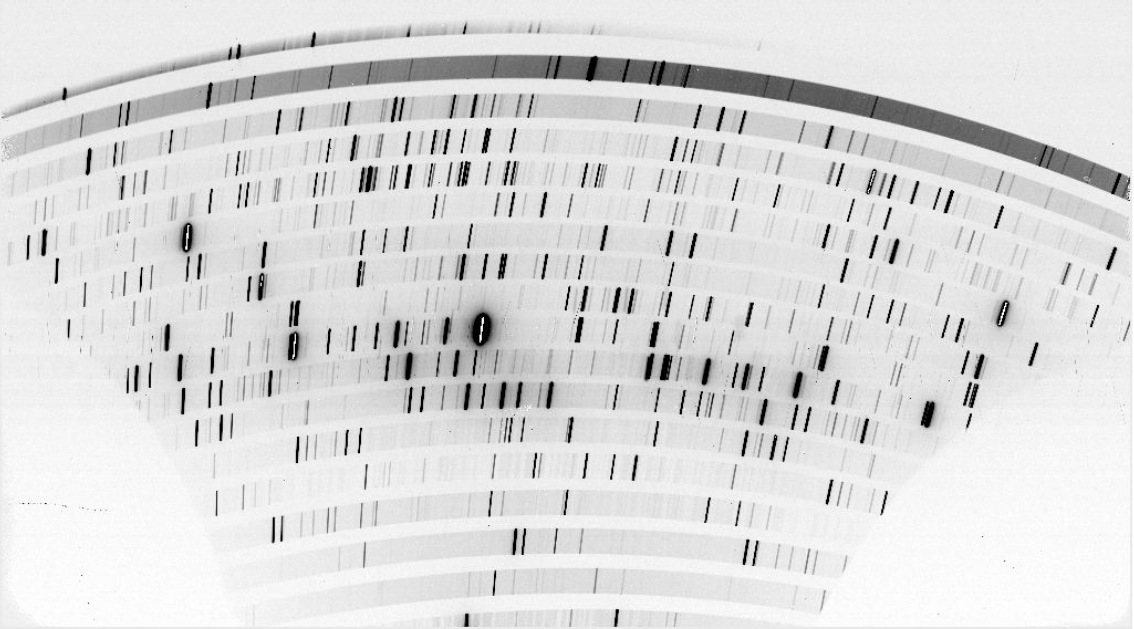}}\\
25&	1020.5	&1046.0	&1076.7&\\
24&	1062.0	&1089.6	&1122.9&\\
23&	1106.6	&1137.0	&1173.1&\\
22&	1155.2	&1188.6	&1228.0&\\
21&	1208.2	&1245.2	&1288.5&\\
20&	1266.5	&1307.5	&1355.2&\\
19&	1330.3	&1376.3	&1429.4&\\
18&	1400.8	&1452.8	&1511.5&\\
17&	1479.5	&1538.2	&1604.0&\\
16&	1567.1	&1634.4	&1708.7&\\
15&	1667.8	&1743.3	&1823.3&\\
14&	1785.7	&1867.9	&1952.8&\\
13&	1922.6	&2011.5	&2102.0&\\
12&	2082.9	&2179.3	&2275.6&\\
11&	2272.3	&2377.28&	2480.7&\\
\hline

  \end{tabular}
\end{table*}

\subsection{image quality and spectral resolution}

In terms of image quality, spectral resolution and sampling, the three arms of
X-shooter perform fully within specifications\footnote{The original
  specifications for the resolution (R=$\lambda$/$\Delta\lambda$) with a 1\arcsec~slit were: R$>$5000 for the UVB arm, 
R$>$7500 for the VIS arm and 
R$>$ 4800 for the NIR arm with a sampling of the line spread function
$>$5 pixels. For a 0.6\arcsec~slit, the specifications were: R$>$7600 for the UVB arm, 
R$>$11500 for the VIS arm and 
R$>$ 7000 for the NIR arm with a sampling of the line spread function
$>$3 pixels.}. Resolution
and sampling as a function of slit width are given in Table~\ref{resol}.

\begin{table*}
\caption{\label{resol} Measured resolution and sampling as a function of slit width.}
\begin{center}
\begin{tabular}{cccc cccc ccc}
\hline\hline
\multicolumn{3}{c}{UVB} & & \multicolumn{3}{c}{VIS}
&&\multicolumn{3}{c}{NIR} \\
\hline
Slit width & Resolution & Sampling&& Slit width & Resolution &
Sampling&& Slit width & Resolution & Sampling\\

(\arcsec) & $(\lambda/\delta\lambda$)&(pix/FWHM)&&(\arcsec) &
$(\lambda/\delta\lambda$)&(pix/FWHM)&&(\arcsec) &
$(\lambda/\delta\lambda$)&(pix/FWHM)\\
\hline
0.5&9100&3.5&&0.4&17400&3.0&&0.4&11300&2.0\\
0.8&6300&5.2&&0.7&11000&4.8&&0.6&8100&2.8\\
1.0&5100&6.3&&0.9&8800&6.0&&0.9&5600&4.0\\
1.3&4000&8.1&&1.2&6700&7.9&&1.2&4300&5.3\\
1.6&3300&9.9&&1.5&5400&9.7&&1.5&3500&6.6\\

IFU&7900&4.1&&IFU&12600&4.2&&IFU&8100&2.8\\
\hline
\end{tabular}
\end{center}
\end{table*}

\subsection{Efficiency} 

\begin{figure}
\begin{center}
\includegraphics[height=8cm,angle=90,trim =20mm 0mm 20mm 0mm, clip]{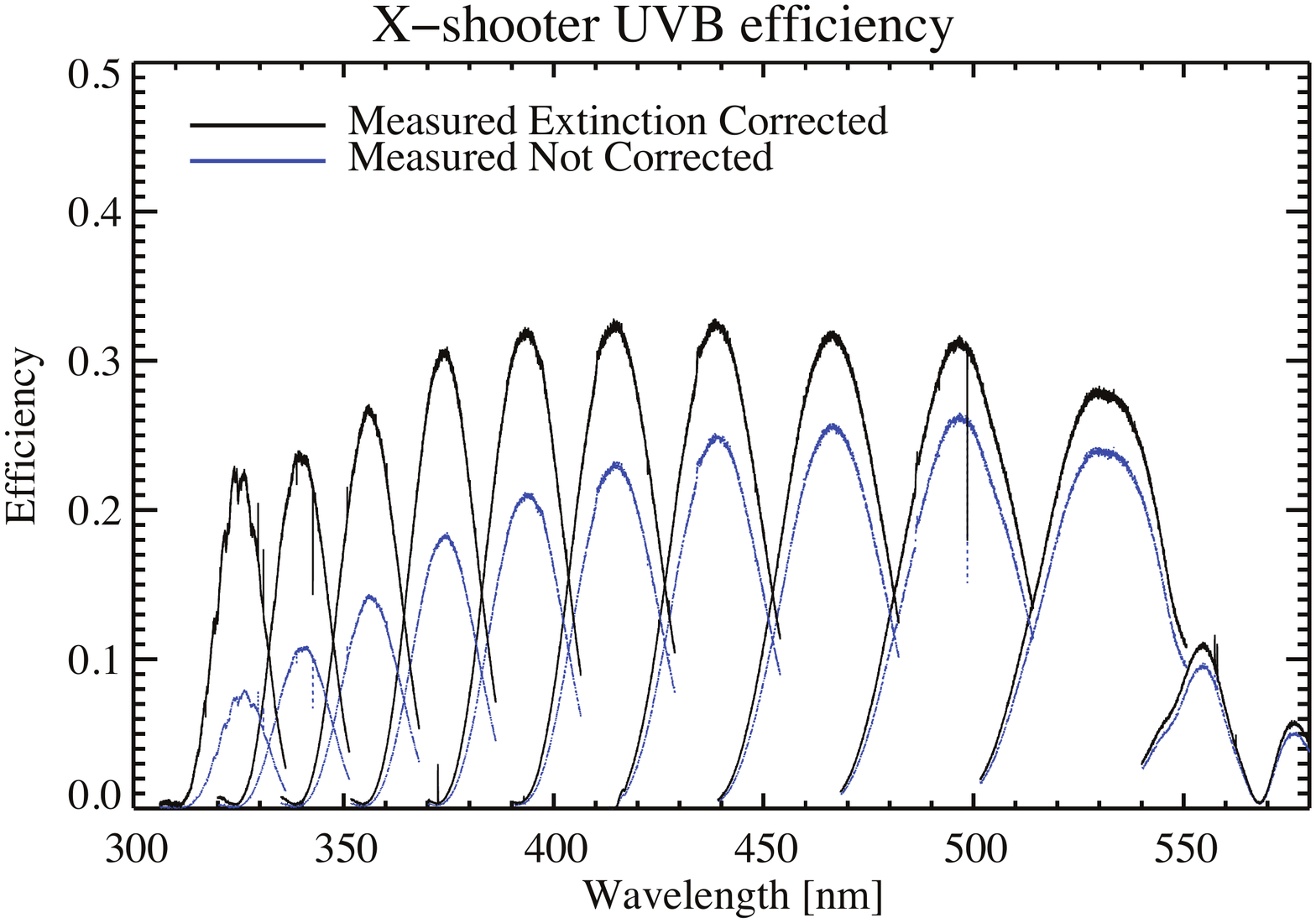} 
\includegraphics[height=8cm,angle=90,trim =20mm 0mm 20mm 0mm, clip]{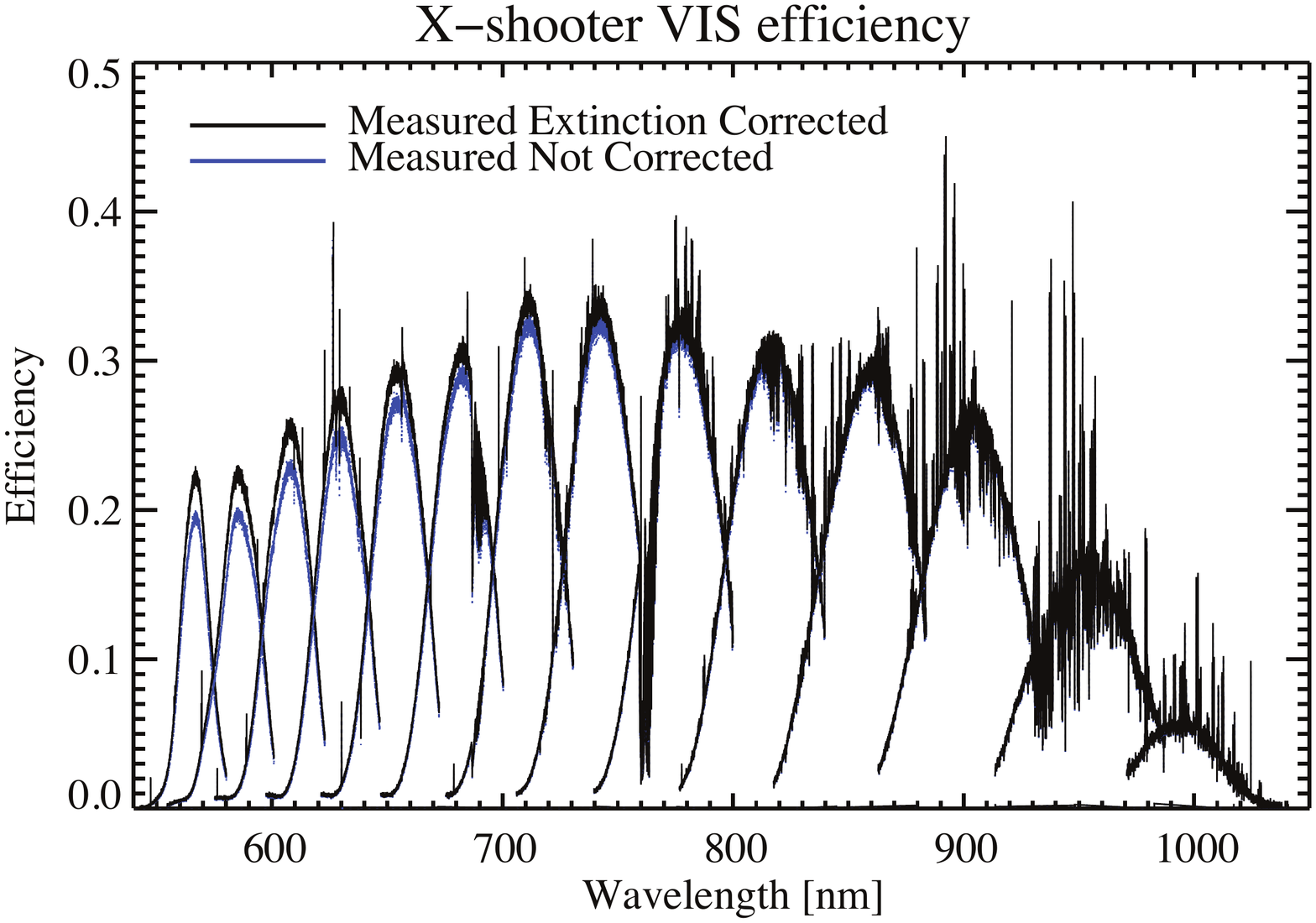} 
\includegraphics[height=8cm,angle=90,trim =20mm 0mm 20mm 0mm, clip]{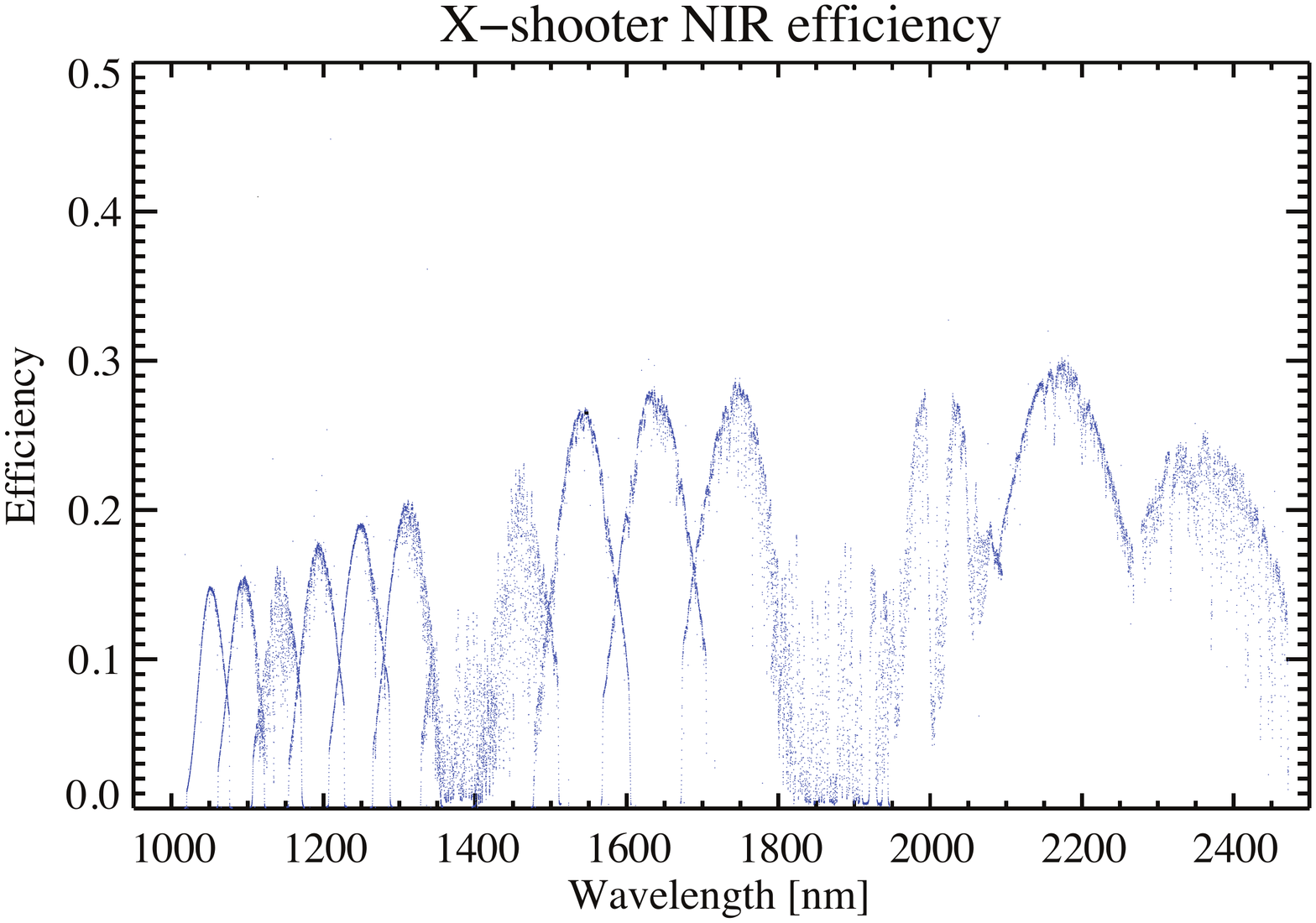}
\end{center}
\caption{\label{efficiency} The overall throughput of X-shooter, including
 telescope, as measured during the last commissioning run in
 september 2009 using spectro-photometric standard BD+17~4708. Blue
 curves represent raw measurements and black curves show the
 efficiency corrected for atmospheric extinction (UVB and VIS only).}
\end{figure}

Thanks to the very high efficiency of the two dichroics splitting
light between the three arms (see Fig.~\ref{dic_fig}) and the careful
optimization of each arm, the resulting overall throughput of
X-shooter is very high. The efficiency for each order as measured
during the last commissioning run using spectro-photometric standard
BD+17~4708 is given in Fig.~\ref{efficiency}. Taking orders
individually (i.e. not combining signal from adjacent orders), the
total efficiency (including telescope and detectors) peaks at 33\%,
34\% and 31\% for the UVB, VIS and NIR arm, respectively.

The overall efficiency is essentially as predicted by multiplying
efficiencies of the different optical elements and detectors except
for the J band where it is $\sim$30\% below our predictions, due to
losses that can only be partly explained by scattering in the ZnSe
cross-disperser prisms.

\subsection{Stability}

Being mounted at the Cassegrain focus, X-shooter is subject to a
changing gravity vector, hence instrument flexure has to be kept under
control. 

This splits into two components: 
(i) 
flexure within each spectrograph (i.e. after the slit) that mainly
affects the quality of the wavelength calibration and the sky
subtraction (see Sect.~\ref{spec_flex});
 (ii) flexure of the
instrument backbone (i.e. before the slit) that affects the relative
alignment of the three spectrographs 
(see Sect.~\ref{bb_flex}).

\subsubsection{Spectrograph flexures}\label{spec_flex}

\begin{figure*}[htpb]
\begin{center}
\subfigure[UVB]{\includegraphics[trim =15mm 0mm 15mm 70mm, clip, height=6.7cm]{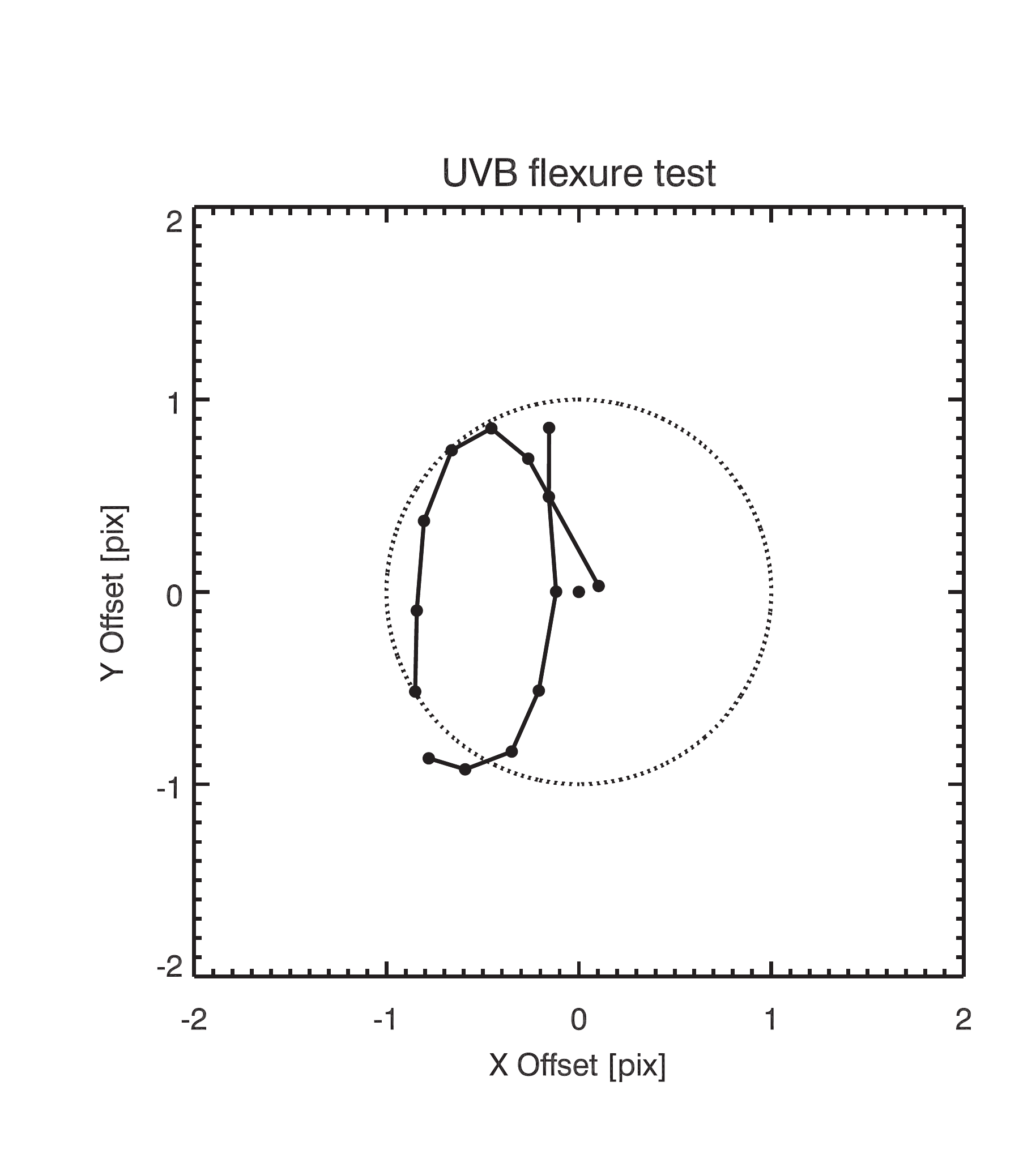} }
\subfigure[VIS]{ \includegraphics[trim =15mm 0mm 15mm 70mm, clip, height=6.7cm]{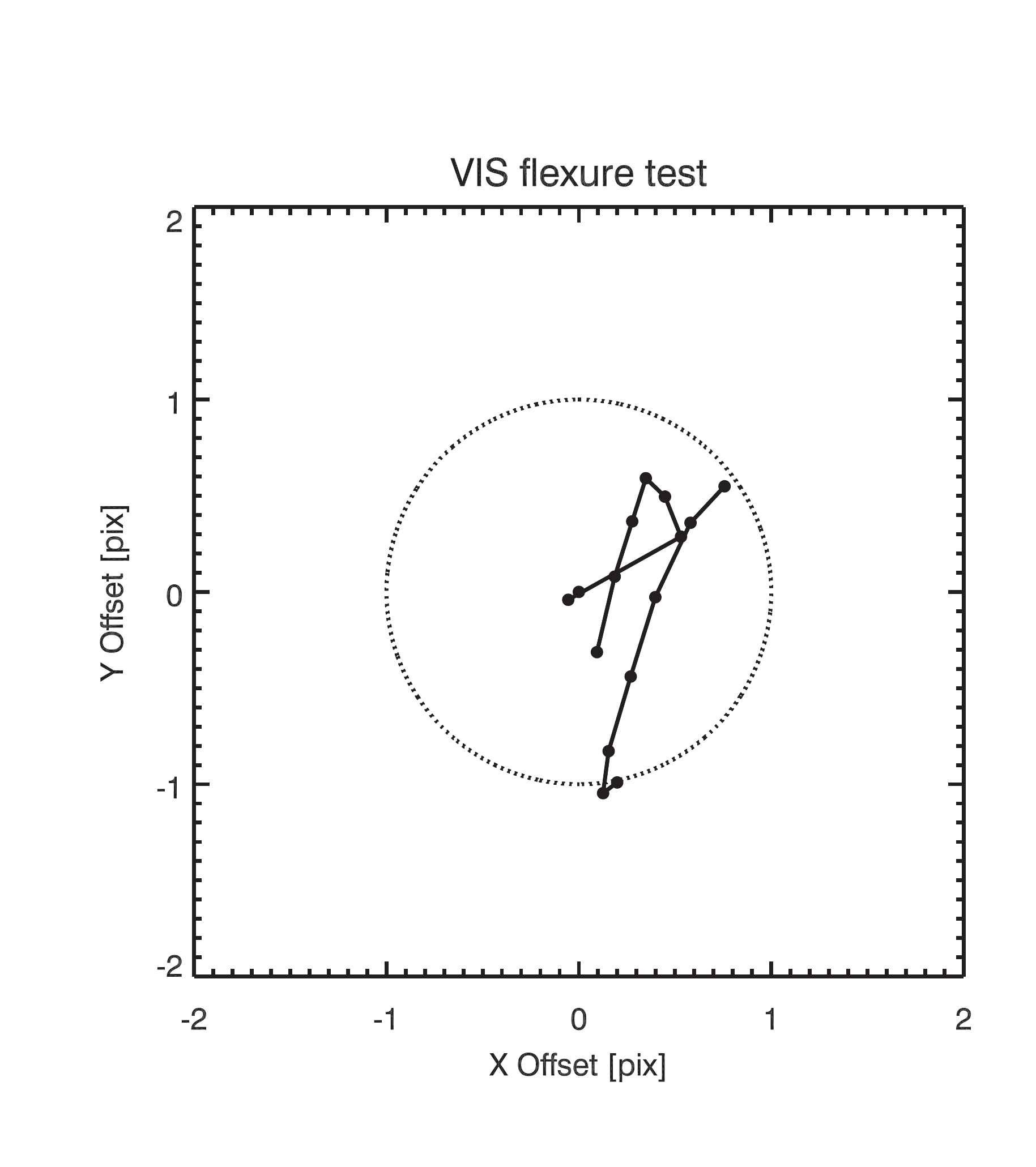}}
\subfigure[NIR]{\includegraphics[trim =15mm 0mm 15mm 70mm, clip, height=6.7cm]{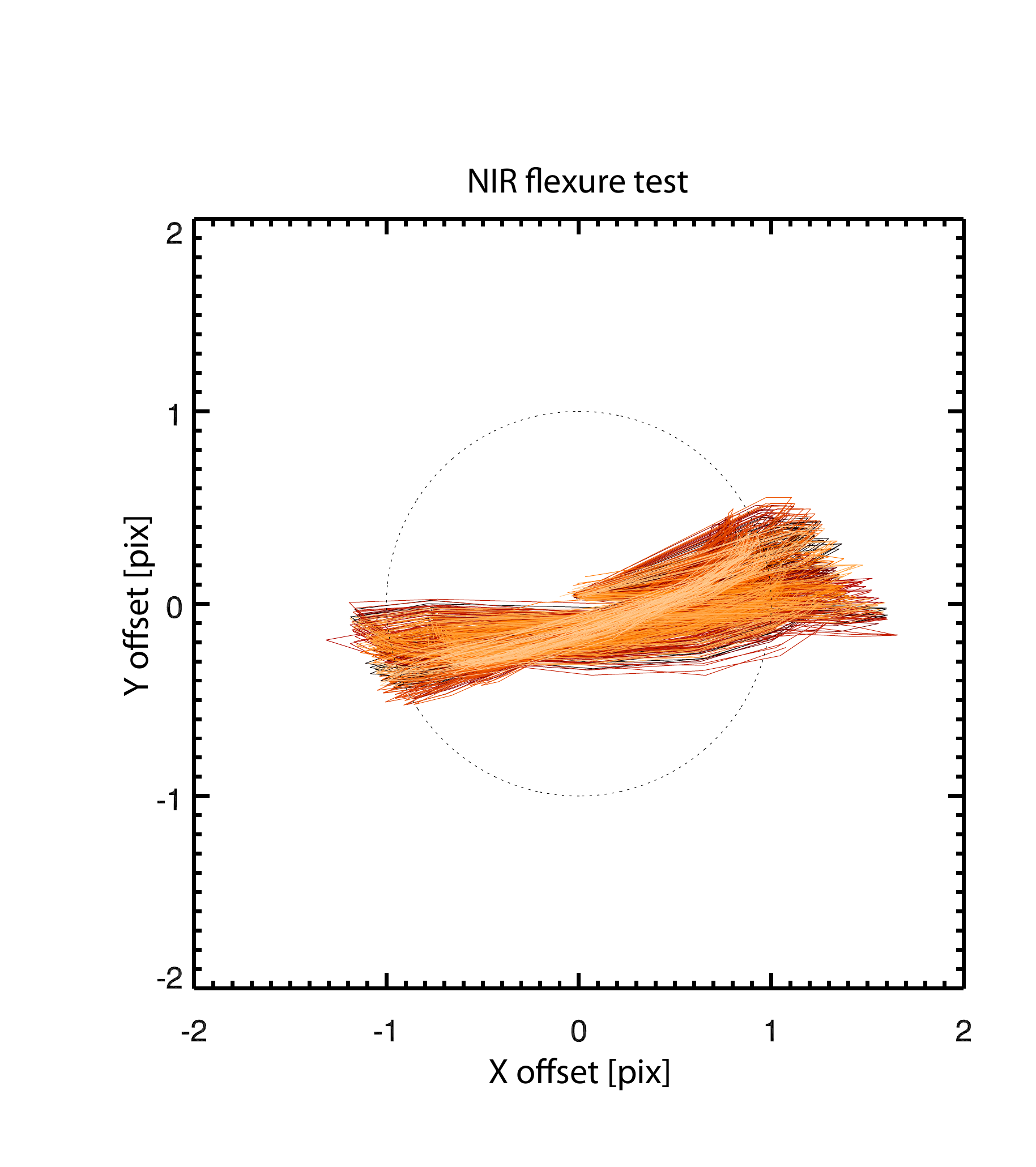}  }
\end{center}
\caption{\label{fig_spec_flex} Measured image position for a full
  rotation of the instrument at a zenith distance of 60\degr\, taking
  as a reference (0,0) the position measured at zenith. X and Y scales
  are in pixels. Pixel scale is $\sim$0.15\arcsec/pix for UVB and VIS,
  and $\sim$0.2\arcsec/pix for the NIR arm. The dotted circle shows
  limits set by our flexure specifications: $\pm 1$ pixel from zenith
  position. In panel (a) and (b), corresponding to the UVB and the
  VIS arm, the curve shows the global motion of the spectral format
  obtained by averaging measurements obtained with five bright
  calibration lines. In panel (c) corresponding to the NIR arm, the
  track followed by many individual calibration lines during the
  rotation of the instrument is shown.}
\end{figure*}

Changes in the spectral format with position have been analyzed in
detail during integration and testing in Garching and further checked
during commissioning in Paranal. Performance at the telescope is
shown in Fig.~\ref{fig_spec_flex}. 

For the UVB and the VIS arm, the image motion measured throughout the
whole detector using many calibration lines is identical, meaning that
flexure induces a simple rigid shift of the spectral format. The
amplitude of this displacement with respect to the position at zenith
for a full rotation of the instrument at a zenith distance of
60\degr\, is $\lesssim$1.15 pixels in the UVB arm and $\lesssim$ 1.0
pixel in the VIS arm as shown in panel (a) and (b) of
Fig.~\ref{fig_spec_flex}. No variation of image quality is measured
for those two arms.

Concerning the NIR arm, the flexure behavior is more complex as
illustrated in panel (c) of Fig.~\ref{fig_spec_flex} which shows with
different colors the recorded image motion for various calibration
lines throughout the spectral format.  Relative to their (x, y)
position at zenith, spectral lines move by up to 1.4 pixels. However,
this effect is larger on the edges than in the center of the detector
hinting at an effect of flexure on image scale. This is possibly due
to some small residual motion of the detector which is linked to a
heavy copper bar reaching the LN2 tank to maintain its low operating
temperature (see Sect.~\ref{nir_det}).  This hypothesis is further
supported by measured variations of the spot FWHM by up to $\sim$15\%
(which, however, stays well within image quality specifications).

Note that the amplitude of the image motion discussed here is based on
the extreme case of a full rotation of the instrument at 60\degr\,
zenithal distance which never occurs during normal observations. In
operation, these shifts are generally small and are not compensated
mechanically.

\subsubsection{Active flexure compensation of the backbone}\label{bb_flex}

One of the main challenges with the three arm design is to keep
the three slits staring at the same patch of sky at any position angle and
zenith distance. In order to always guarantee an alignment to
better than 1/10$^{\rm th}$ of the narrowest slit width, X-shooter is equipped
with an Active Flexure Compensation (AFC) system: each arm has a piezo
mounted folding mirror which is adjusted during each acquisition, immediately
after the telescope and the instrument have reached their position for science
observations. Flexure is measured via application of the follwing recipe:

\begin{figure*}
\begin{center}
\subfigure[AFC step 1]{
\includegraphics[height=6.4cm]{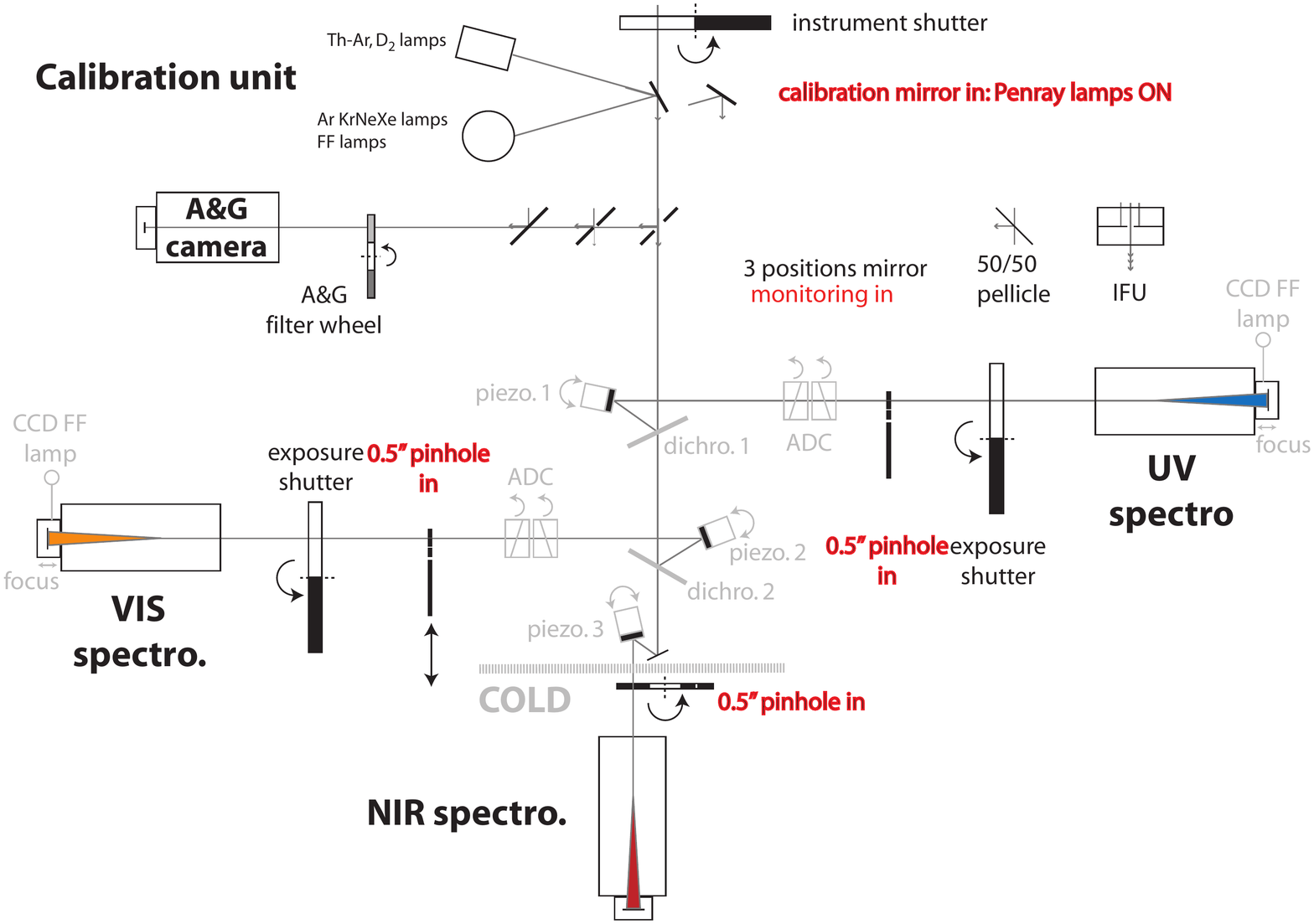}}
\subfigure[AFC step 2]{
\includegraphics[height=6.4cm]{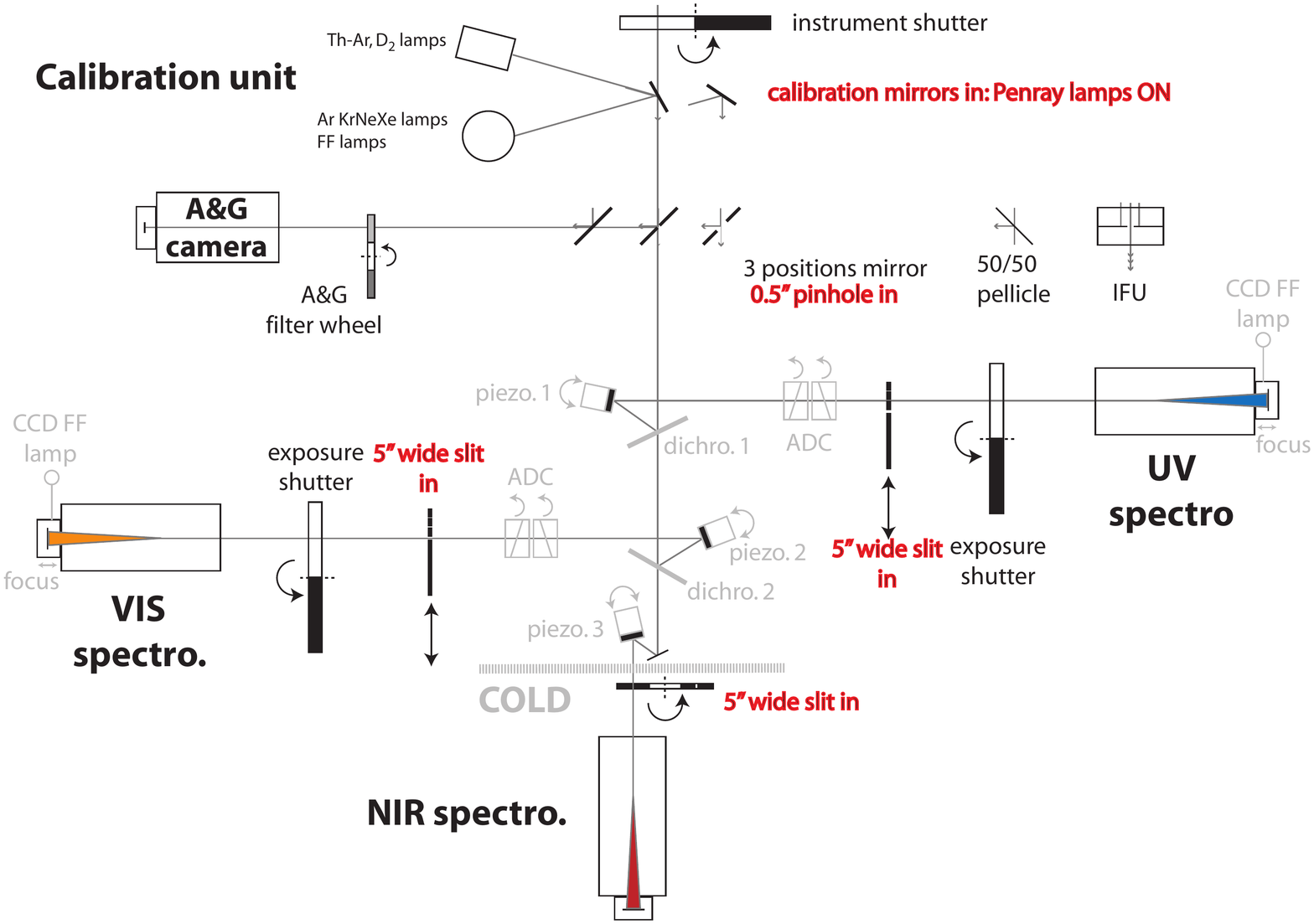}}
\subfigure[AFC step 3]{
\includegraphics[height=6.4cm]{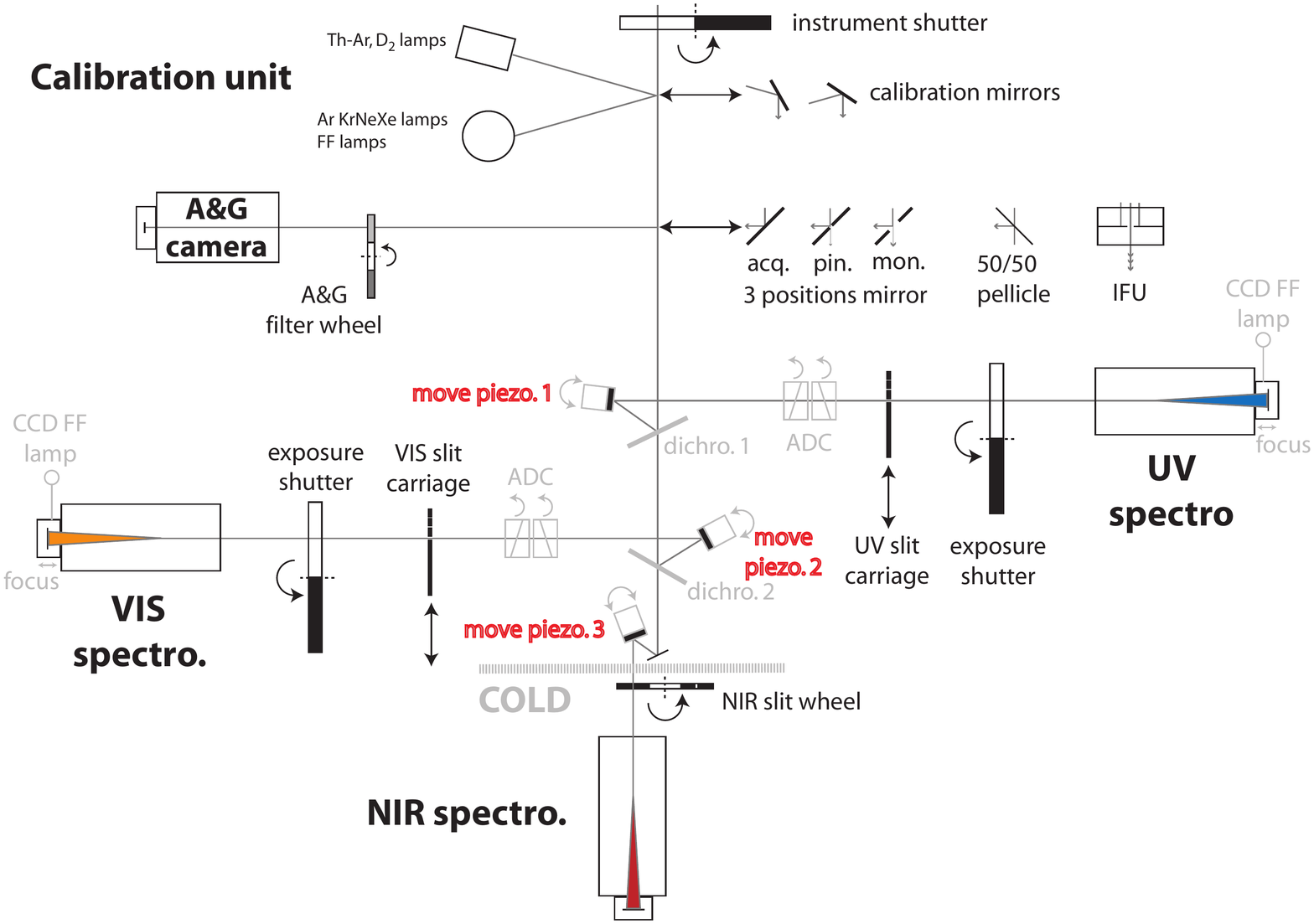}}
\end{center}
\caption{\label{afc}The three steps of the flexure compensation
  procedure. For each step, functions that are moved are highlighted
  in red. In step 1, panel (a), the ArHgNeXe calibration lamps are
  switched on, the 3-position mirror in the A\&G slide is set to the
  wide slot position and the 0.5\arcsec~pinhole is inserted in the
  slit unit of each spectrograph ; in step 2, panel (b), the
  calibration lamp is still on, the 3-position mirror is set to the
  0.5\arcsec~pinhole position and the wide 5\arcsec~slit is inserted
  in each slit unit ; in step 3, panel (c), the piezo mounted folding
  mirrors are moved according to the measurements obtained in step 1
  and 2.}
\end{figure*} 

\begin{enumerate}
\item take simultaneously in the 3 arms an arc spectrum
  through the 0.5\arcsec~pinhole located in the spectrograph slit
  slide/wheel (Fig.~\ref{afc}a);
\item take a spectrum using a reference 0.5\arcsec~pinhole in the
  cassegrain focal plane (see Fig.~\ref{afc}b);
\item measure the displacement between the two frames (at the
  undeviated wavelength of the atmospheric dispersion compensator)
  using a cross-correlation algorithm;
\item send corresponding commands to piezos  (Fig.~\ref{afc}c);
\item repeat steps 2 \& 3 to check convergence.
 \end{enumerate}

\begin{figure}
\begin{center}
\includegraphics[trim =0mm 0mm 10mm 60mm, clip, width=9cm]{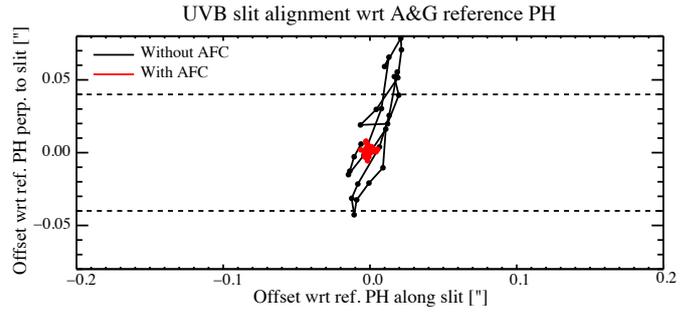} 
\end{center}
\caption{\label{flex_afc} Alignment of the UVB arm 0.5\arcsec~pinhole
with respect to the focal plane reference pinhole with (red) and
without the flexure compensation (black) through a full rotation of
X-shooter at a zenith distance of 60\degr. The AFC allows to
maintain the alignment to $\sim$0.01\arcsec, well within
specification of 1/10th of the narrowest slit width.}
\end{figure} 

 This whole procedure comes at no expenses in terms of overheads since
 it is done in parallel with the telescope active optics setup. It is
 operationally very robust and does not require any user
 interaction. Our measurements show that it reliably maintains the
 alignment of the three slits to better than 0.02\arcsec, as illustrated in
 Fig.~\ref{flex_afc}. As a side product, the first frame of the
 sequence is actually an ``attached'' wavelength calibration that is
 used by the data reduction pipeline to correct the wavelength
 solution for for thermally- and gravity-induced drifts \citep[see][]{modigliani2010}.

\begin{figure}
\begin{center}
\includegraphics[height=9.1cm, angle=90,trim =20mm 0mm 20mm 0mm, clip]{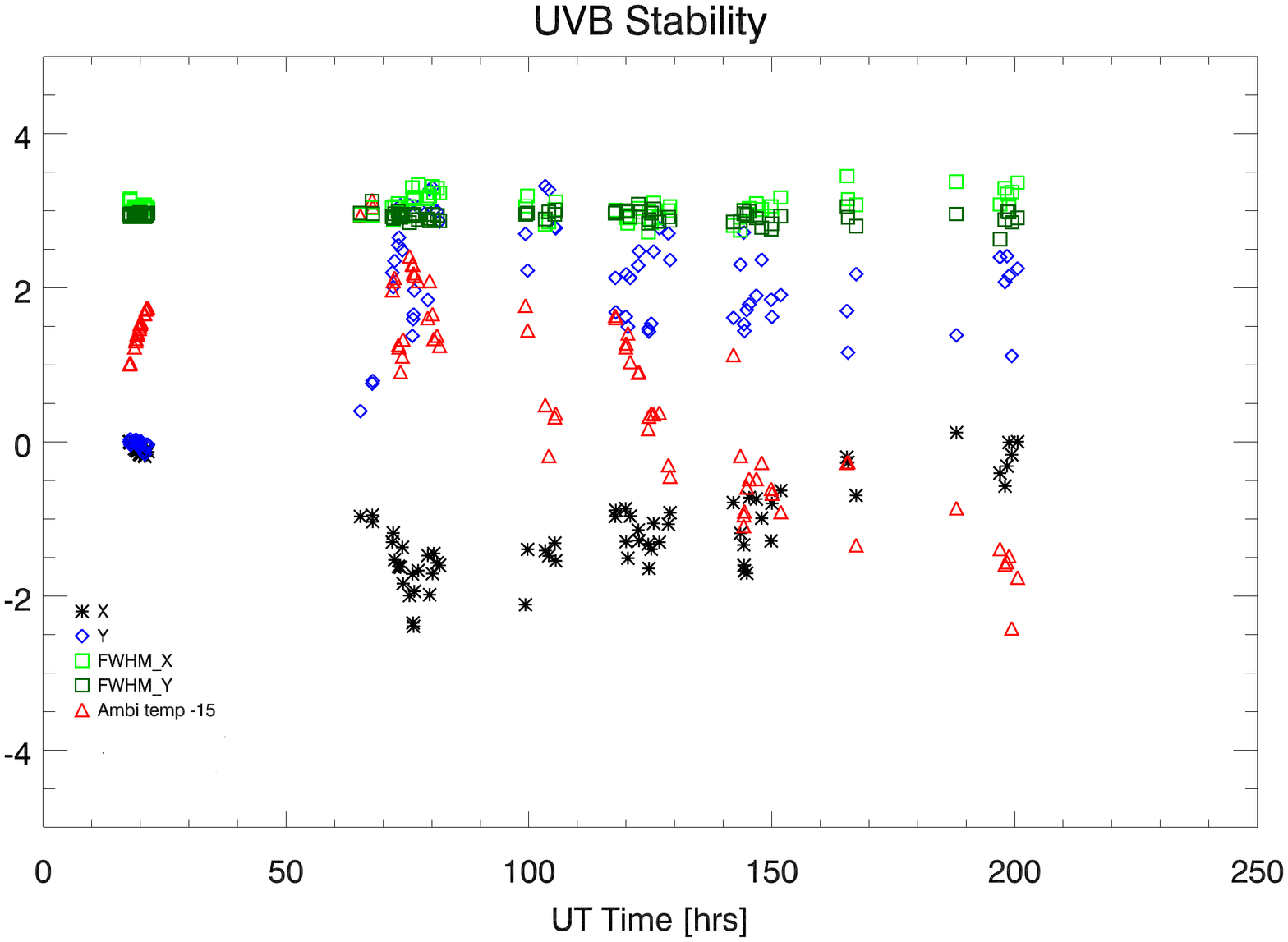}\\
\includegraphics[height=5.5cm,trim =10mm 20mm 10mm 30mm, clip]{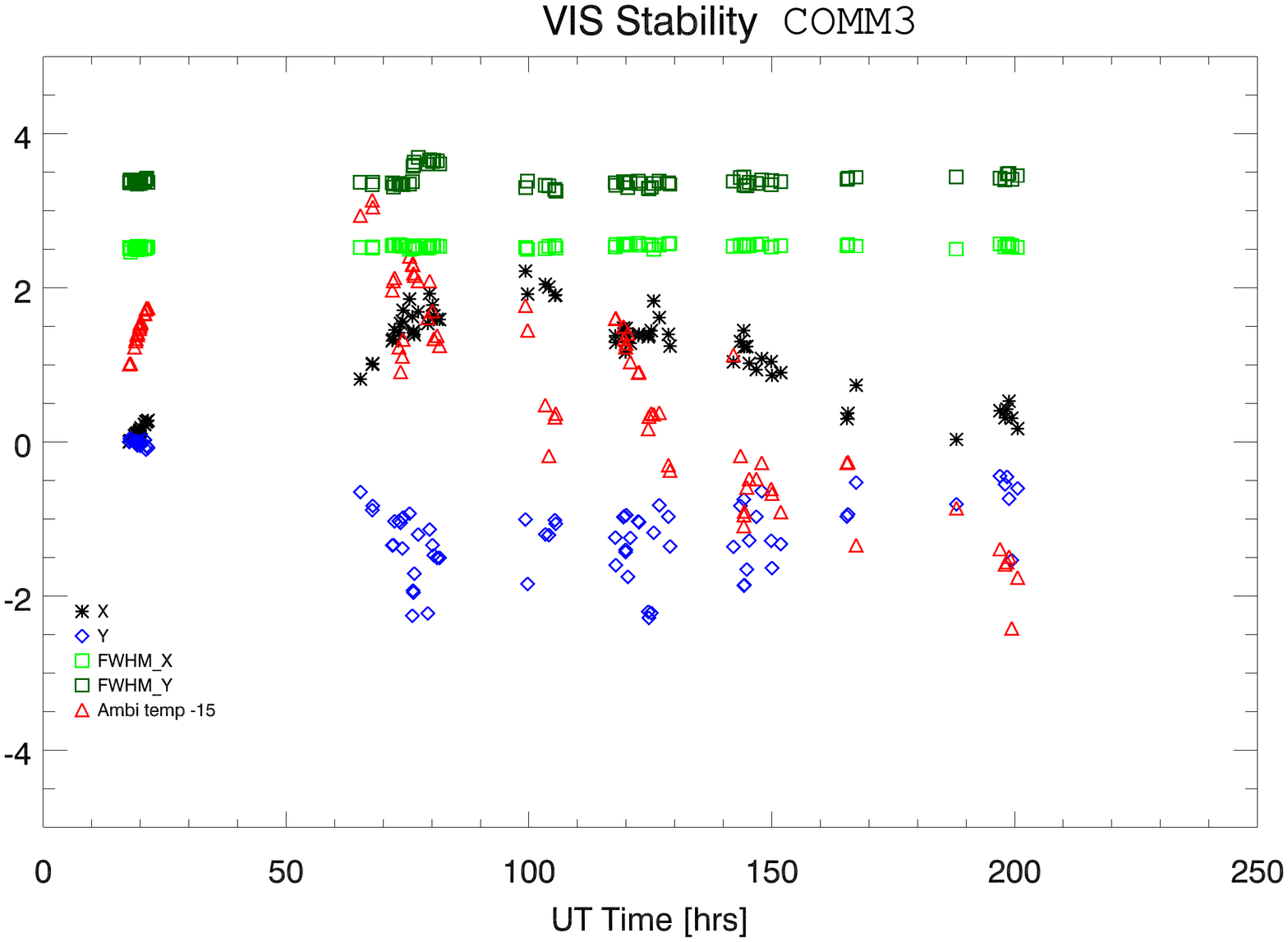}\\
\includegraphics[height=5.5cm,trim =10mm 20mm 10mm 30mm, clip]{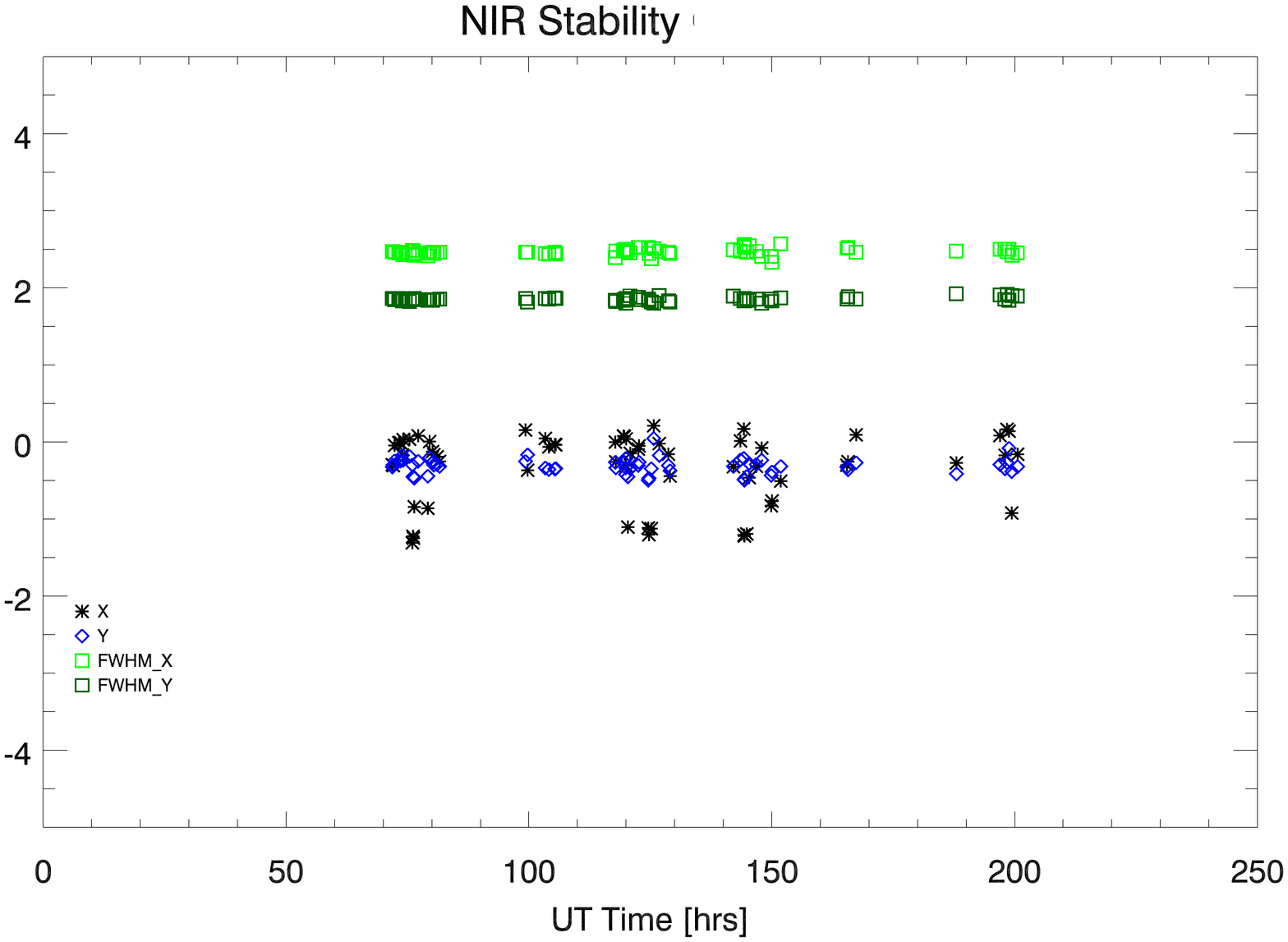}

\end{center}
\caption{\label{stability} Stability of the UVB (top), VIS (center) and
 NIR (bottom) spectrographs over 6 consecutive nights as
 measured during the third commissioning run. The dark and light
 green points show the FWHM in X and Y of a 0.5\arcsec pinhole, the
 black and blue points show the X and Y position of a reference
 calibration line with respect to its position measured at zenith on the first
day and the red points show the ambient temperature (for UVB and VIS
only since it is irrelevant for the temperature controlled NIR arm).}
\end{figure}

\subsection{Near-IR arm background}\label{nirbck}

The near-infrared sky is very dark in between the OH sky emission
lines \citep{maihara93}. In order to reach sky background limited
conditions, keeping the instrument background at the lowest possible
level is therefore of utmost importance. Though a critical aspect of the
performance of any near-IR spectrograph, it is however quite challenging
and requires rigorous design and careful manufacturing (eg. good
baffling, limited number of cable feed-throughs).

The near-IR arm of X-shooter is a remarkably dark instrument: with a
closed slit the measured background is below 0.01 e$^{-}$/s/pix. This
means that assuming a sky background level as measured by
\cite{maihara93}, the instrument background should be a factor
two to three below the sky. In addition, in an exposure of 1800s, the
photon noise from the instrument background would be around 4 e$^{-}$
i.e. less than the read noise of the detector.

\begin{figure}
\begin{center}
\includegraphics[width=9cm]{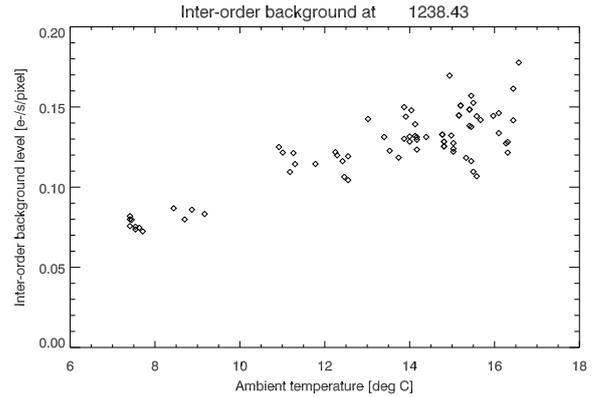} 
\end{center}
\caption{\label{nir_bck} Background level measured in the
  inter-order background region on both sides of order 21 at 1238.64
  nm versus ambient temperature.}
\end{figure}

However, on sky measurements have revealed a background level
estimated from the intensity of the inter-order space in the J- and
the H-band a factor of three to five higher than the sky level. This
stray light flux level rises in proportion to the slit width. A complete
analysis of the background level using all commissioning data spaning
a range of observing conditions further showed that this spurious
background is correlated with ambient temperature as shown in
Fig.~\ref{nir_bck}. This indicates that it is probably caused by K
band radiation that is reflected by the detector into the camera and
then comes back to the detector as a diffuse stray light background.

The immediate consequence is that in the J and H-band the
instrument in its present state is clearly not sky background limited.

\section{Overview and future upgrades}\label{conclusion}

In this paper, we have presented key figures demonstrating X-shooter's
high performance on sky. This unique instrument lives up to
expectations in terms of high throughput, resolution and exquisite image
quality. It is overall a bit less stable than originally targeted but
flexures are nevertheless kept within a very reasonable range
ensuring no significant impact on science performance. The novel
concept developed to compensate for backbone flexures and accurately
maintain the three slits staring at the same patch of sky is very
efficient and robust in operations.

After two ``science verification'' runs\footnote{All science
  verification proposals and data are publicly available at
  http://www.eso.org/sci/activities/vltsv/xshootersv.html} in August
and September 2009, the instrument started regular operation in
period 84, that is on October 1$^{\rm st}$, 2009. In the first four Calls
for Proposals for the VLT in which X-shooter has been offered (ESO
Period 84 to 87), it has been the second most requested of the 14
available instruments, with a ratio requested versus scheduled
observing time of 3.2.  The first year of operations confirms the high
versatility of the instrument and the very broad range of topics
tackled by X-shooter, as anticipated in the original science case
analysis, from stellar astrophysics \citep[see e.g.][]{bragaglia2010,
  schwope2010, rigliaco2011, kaper2011}, cosmic explosions
\citep[e.g.][]{deugarte2010, delia2010} to the high redshift Universe
\citep[e.g.][]{dessauges2010, fynbo2010, christensen2010}.

Several ideas to further improve the performance of X-shooter or add
new functionalities have been proposed. The most advanced one concerns
the reduction of the J- and H-band background. As explained in
Sect.~\ref{nirbck}, due to scattered light from the very bright
thermal background dominated K-band orders, sky limited observations
in the J- and H-band are currently not possible. There was a choice of
either simply baffling the longest wavelength orders or placing a cold
filter in front of some of the slits. The second option was chosen and
two new slits (0.6\arcsec and 0.9\arcsec wide) equipped with a short
pass filter blocking the spectral range above 2$\mu$m will be fitted
into the NIR spectrograph slit wheel. Blocking the K-band radiation
should restore the expected low background at the expense of the
spectral range for those two new slits. In order to allow the
acquisition of faint red sources --- critical in the case of some GRB
observations for instance ---, the feasibility of adding a near-IR
channel to the acquisition system is being investigated. In addition,
the possibility to add spectro-polarimetric capabilities and the idea
of replacing the piezo mounts of the folding mirrors to allow
(counter)nodding in the UVB and VIS arms (i.e. nodding along the slit
in the NIR arm while keeping the target at the same position in the
UVB and VIS arm) have also been proposed.

\section{The spectrum of the QSO B1422+231.}\label{example}
\begin{figure*}
 \begin{center}
\includegraphics[width=14cm]{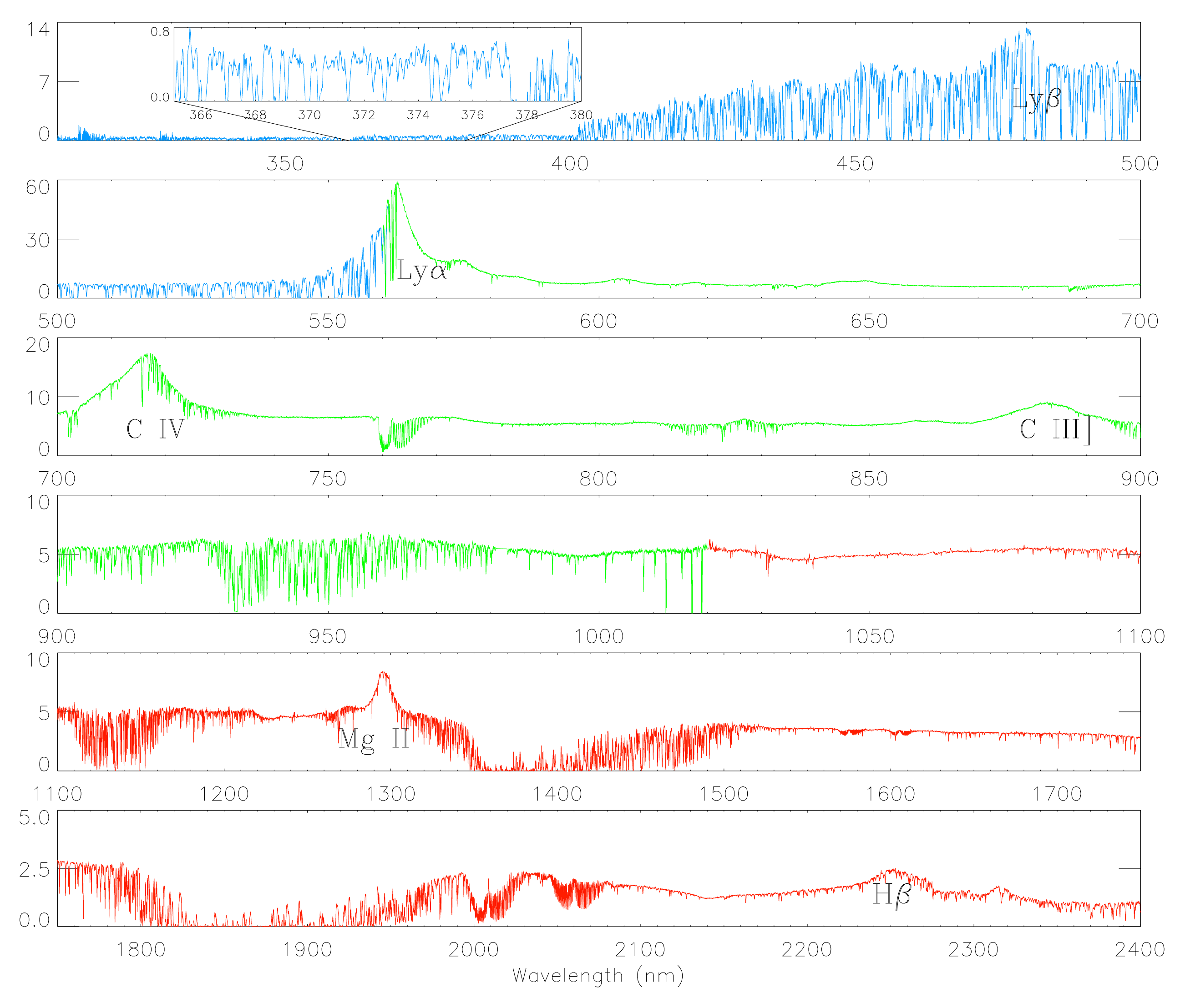}
\end{center}
\caption{ \label{figB1422} Spectrum of the lensed QSO B1422+231. The
  blue part of the spectrum shows the UVB, the green the VIS, and the
  red the NIR data. No
  correction for telluric lines has been applied.}
\end{figure*}

Finally, as an example of the observing capability of the instrument,
we show in Fig.~\ref{figB1422} the reduced X-shooter spectrum of
B1422+231 (z =3.62). This QSO is sufficiently bright (V\,$\sim~$16.5
mag) due to gravitational lensing to be observed at medium-high
resolution with high signal-to-noise in about 1 hour.  These data were
obtained during the commissioning of the instrument in its full
configuration in March 2009 to secure a template spectrum of a QSO
over the full spectral range. The spectrum covers the wavelength
interval 70-520 nm in the rest frame of the QSO. The integration time was 4800s
split over 4$\times$1200s exposures. The brightest two of the lensed
QSO images, separated by 0.5\arcsec, were aligned along the slit and
the extracted spectrum refers to the sum of the two. The reduction was
carried out with the standard X-shooter data reduction package. The
final signal-to-noise ratio is between 50 and 100 over most of the
spectrum.  The spectral resolution is 6,200, 11,000 and 8,100 in the
UVB, VIS and NIR spectral instrument arms respectively (on average two
pixel sampling of the resolution element). The wavelength scale is
calibrated to an accuracy of better than 2 km/s in the VIS arm
and better than 4 km/s in the UVB and NIR arms, as verified on sky
emission lines. The spectrum shown in Fig.~\ref{figB1422} has been
corrected for relative spectral response of the instrument via a
standard star to a 5\% accuracy. Since the night was not photometric,
an accurate absolute flux scale could not be established.

\begin{figure}
 \begin{center}
\includegraphics[width=9cm,trim =0mm 0mm 0mm 70mm, clip]{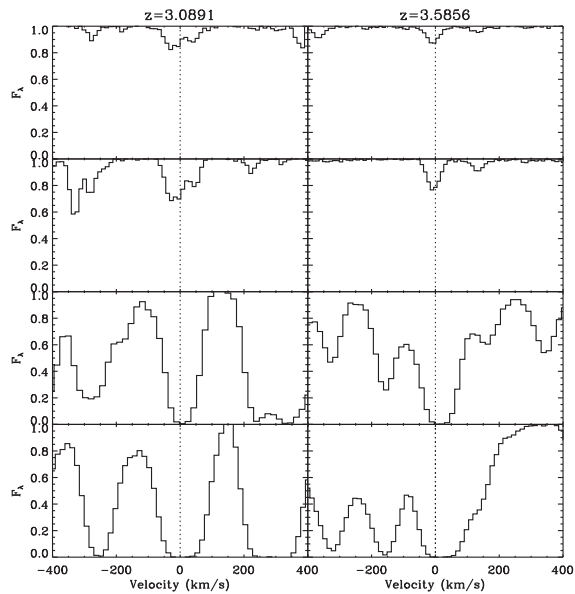}
\end{center}
\caption{ \label{absorbers} 
The C~{\sc{iv}}~$\lambda$1548\,\AA\,and the
C~{\sc{iv}}~$\lambda$1550\,\AA\,(top) lines of two high redshift metal
systems in the line of sight to the QSO B1422+231 are shown in these
normalized plots together with the corresponding Ly$\alpha$ and
Ly$\beta$ absorptions. The data compare well with the observations of
these systems with HIRES at Keck at three times the resolving power
and from longer integration times published by \cite{songaila96}. With
the X-shooter, it was possible to observe all of these lines in
parallel exposures within the blue and visual arm. The corresponding
Mg II absorptions of the same systems fall in the NIR arm of the
instrument.
}
\end{figure}

The quality of the X-shooter spectra can be also judged from a
comparison with the best high resolution data of this QSO in the
literature. \cite{songaila96} discuss the metal absorption systems in
the line of sight to this QSO from Keck HIRES spectra at resolution
36000 (total exposure time 8.3 hrs). Fig.~\ref{absorbers} shows two C
IV metal systems for which the corresponding Ly$\alpha$ and
Ly$\beta$ are also observed in the same exposure. A comparison
with the corresponding tracings in Fig.~2 of \cite{songaila96} shows
that all the components of the metal systems are detected although the
resolution of the X-shooter is nominally a factor of 3 lower and the
exposure time a factor of 4 shorter.

This is the first time that a QSO is observed over such a wide
spectral range in a single observation. With a standard optical or
infrared spectrograph only limited regions of the spectrum could be
studied with a single exposure, with the potential risk of introducing
errors in the final compilation of data taken at different times and
under different weather conditions. With X-shooter data it is now
possible to circumvent these obstacles.

\begin{acknowledgements}

  The X-shooter project acknowledges financial support from the EU
  Descartes prize 2002 "Solving the gamma-ray burst riddle: the
  universe's biggest explosions" (PI Ed van den Heuvel), the Carlsberg
  foundation, the Netherlands Research School for Astronomy NOVA, and
  the Netherlands Organization for Scientific Research NWO. The design
  and implementation of the control software of X-Shooter was carried
  out in the framework of a PRIN project of the MIUR (Italian Ministry
  of Education, University and Research). We thank J. Prochaska for
  discussions on the optimal approach to the data reduction in the
  early phases of the project.

\end{acknowledgements}

\bibliographystyle{aa} 
\bibliography{all} 
\end{document}